\RequirePackage{fix-cm}
\documentclass[twocolumn,epjc3]{svjour3}  
\makeatletter
\renewcommand{\makeheadbox}{}
\makeatother
\smartqed  
\RequirePackage{graphicx}
\RequirePackage[caption=false,font=normalsize,labelfont=sf,textfont=sf]{subfig}
\usepackage{tikz}
\usepackage{microtype}
\usepackage{graphicx} 
\usepackage{caption}  
\usepackage{booktabs}  
\usepackage{array}     
\usepackage{multirow}  
\usepackage{tabularx}
\usepackage{array}
\usetikzlibrary{shapes.geometric, arrows.meta, positioning}
\usetikzlibrary{intersections, backgrounds, decorations.pathreplacing}
\usetikzlibrary{pgfplots.groupplots, matrix, fit}
\usetikzlibrary{arrows.meta, positioning, calc}
\usepackage{stfloats}    
\usepackage{textcomp}    
\usepackage{orcidlink}
\usepackage{caption}
\usepackage{bbding}
\usepackage{pifont}
\usepackage{xcolor}
\usepackage{colortbl}

\usepackage[table]{xcolor}
\usepackage{amssymb}

\journalname{International Journal of Information Security}
\begin{document}

\title{eBPF-Based Cybersecurity Mechanisms: A Systematic Literature Review}


\author{Stamatios Kostopoulos \orcidlink{0009-0004-8388-0134}\thanksref{e1}\Envelope
        \and
        Panagiotis Tsakonas \orcidlink{0009-0003-3336-6885}\thanksref{e2}
        \and
        Evangelos K. Markakis \orcidlink{0000-0003-0959-598X}\thanksref{e3}
}

\thankstext{e1}{Corresponding author: Stamatios Kostopoulos; s.kostopoulos@pasiphae.eu}
\thankstext{e2}{Panagiotis Tsakonas: p.tsakonas@pasiphae.eu}
\thankstext{e3}{Evangelos K. Markakis: emarkakis@hmu.gr}



\institute{Department of Electrical \& Computer Engineering,\\
           Hellenic Mediterranean University,\\
           Heraklion, Greece
}

\date{}

\maketitle

\begin{abstract}
Extended Berkeley Packet Filter (eBPF) has emerged as a kernel-level programmable framework enabling dynamic security enforcement in modern operating systems. While eBPF's potential for cybersecurity applications has attracted significant research attention, existing work remains fragmented across disparate domains, evaluation methodologies, and deployment contexts. This systematic literature review applies PRISMA methodology to identify, categorize, and synthesize peer-reviewed research on eBPF-based cybersecurity mechanisms. Following structured screening of 3735 records from six databases, 54 primary studies published between 2018-2026 were analyzed and organized into a seven-domain taxonomy spanning DDoS mitigation, intrusion detection, Internet-of-Things (IoT) security, container security, microservice protection, networking, and security tools and frameworks. Analysis reveals that eBPF enables low-overhead security enforcement (median 2.4\% overhead [1.1 - 8.6\%] of the average CPU usage, ranging from negligible nanosecond-scale costs for infrequently used hooks to higher 10 - 20\% CPU percentages for kernel hot paths) with high detection accuracy (94-99\%) across domains, particularly excelling in kernel-level monitoring, real-time packet processing, and cloud-native workload protection. However, significant challenges persist: verifier-imposed constraints limit algorithm complexity, 85.1\% (46/54) of studies require low-level programming expertise, kernel version fragmentation hinders portability, and 96.2\% (52/54) of research fails to address eBPF's own security vulnerabilities. The review identifies critical research gaps in multi-tenant isolation, adversarial machine learning (ML) robustness, production validation, and standardized evaluation frameworks. By consolidating fragmented knowledge and highlighting architectural trade-offs between safety and expressiveness, this work provides a foundation for next-generation eBPF security systems and outlines actionable directions for kernel programmability research.

\keywords{eBPF \and Cybersecurity \and Attack Detection \and Network Security \and Systematic Literature Review}
\end{abstract}

\section{Introduction}
\label{intro}
eBPF has emerged as a kernel-level programmable framework that enables safe and efficient execution of user-defined code within operating system kernels. By combining in-kernel verification, just-in-time compilation, and event-driven execution, eBPF supports high-performance networking, fine-grained observability, and dynamic system instrumentation across modern computing environments, including cloud-native infrastructures, containerized platforms, and IoT deployments.

These capabilities have motivated a growing body of research exploring eBPF for cybersecurity purposes, such as network intrusion detection, runtime workload protection, kernel integrity monitoring, and distributed denial-of-service mitigation. Prior studies demonstrate that in-kernel programmability can enable low-latency detection and enforcement while reducing reliance on user-space monitoring pipelines. However, existing research remains fragmented across heterogeneous application domains, evaluation methodologies, and deployment assumptions, making it difficult to derive a unified understanding of eBPF’s effectiveness, limitations, and architectural implications for modern security systems.

\subsection{Positioning Relative to Existing Surveys}

As seen in Table~\ref{tab:survey_comparison}, although several surveys \cite{survey1}, \cite{survey2}, \cite{survey3}, \cite{survey4}, \cite{survey5} discuss eBPF in the contexts of observability, networking, or performance optimization, a comprehensive synthesis focused specifically on cybersecurity applications and defensive mechanisms is currently lacking. In particular, there is limited comparative analysis of security use-cases, absence of standardized benchmarking practices, and insufficient exploration of emerging directions such as explainable detection mechanisms and AI-assisted in-kernel security analytics. These gaps hinder the systematic design and evaluation of next-generation eBPF-enabled security architectures.

More specifically, the only survey that was fully security-based was that of Her et al. at ~\cite{survey2}, but with the important difference of it providing a targeted empirical evaluation of four prominent eBPF and Kubernetes-based cloud security tools. In contrast, this SLR synthesizes several security-broad studies and goes beyond tool comparison to evaluation methodologies, threat model analysis, deployment contexts, architectural limitations and emerging research challenges, offering a more comprehensive perspective on the role of eBPF in cybersecurity.

\begin{table}[!ht]
\centering
\caption{Comparison with Existing eBPF Surveys}
\label{tab:survey_comparison}
\scriptsize
\setlength{\tabcolsep}{3pt}
\begin{tabularx}{\columnwidth}{@{} X c >{\raggedright\arraybackslash}p{2.2cm} c c @{}}
\toprule
\textbf{Survey} & \textbf{Year} & \textbf{Primary Focus} & \textbf{Sec.} & \textbf{Method} \\
\midrule
Soldani \cite{survey1} & 2023 & 5G/6G Observability & Partial$^a$ & Narrative \\
Her \cite{survey2} & 2025 & Cloud-native Tools & Yes & Empirical \\
Deokar \cite{survey3} & 2024 & Dev. Challenges & No & Empirical \\
Song \& Li \cite{survey4} & 2024 & General Overview & No & Narrative \\
Sedghpour \cite{survey5} & 2022 & Service Mesh & Partial$^b$ & Narrative \\
\midrule
\textbf{This work} & \textbf{2026} & \textbf{Cybersecurity} & \textbf{Full} & \textbf{SLR} \\
\bottomrule
\end{tabularx}
\vspace{2pt}
\begin{flushleft}
\scriptsize
$^a$Covers network isolation and privacy, but not comprehensive threat analysis.\\
$^b$Discusses service-to-service authentication and encryption only.
\end{flushleft}
\end{table}

Key differentiators of this systematic literature review include:

\textbf{Security-Exclusive Focus:} While prior surveys mention security as one application among many (networking, observability, performance), this work exclusively examines cybersecurity mechanisms, enabling deeper analysis of threat models, attack scenarios, and defense architectures that remain underexplored in general eBPF surveys.

\textbf{Systematic Methodology:} Unlike narrative or empirical surveys, this work employs PRISMA \cite{page2021prisma} guidelines with formal inclusion/exclusion criteria and structured screening phases. Reproducible search protocols were executed across six databases, ensuring methodological rigor and transparency absent from previous reviews.

\textbf{Comprehensive Threat Taxonomy:} Prior work does not systematically categorize eBPF security applications across threat types (DDoS, intrusion detection, malware detection, privilege escalation) or deployment contexts (IoT, containers, microservices, networking). This review provides the first structured taxonomy spanning seven security domains with explicit threat-mechanism-efficacy mappings.

\textbf{Limitations and Trade-offs Analysis:} Existing surveys emphasize eBPF's capabilities 
but provide limited analysis of verifier constraints, kernel version fragmentation, or fundamental 
safety-expressiveness trade-offs. This work synthesizes architectural limitations across 54 studies, identifying critical gaps in multi-tenant isolation, where mechanisms for secure sharing of eBPF infrastructure remains underexplored; meta-security, referring to the protection of the eBPF subsystem itself (such as the verifier, JIT compiler and runtime) against eBPF bypasses; and production validation, where most solutions lack real-world evaluation.

\textbf{Temporal Coverage:} With studies through 2026, this review captures recent developments in ML-eBPF integration, zero-trust architectures, and cloud-native security that postdate earlier surveys.

To address this issue this research complies with PRISMA’s guidelines, ensuring substantial transparency that includes core components of literature review and meta-analysis, such as flow diagrams with search criteria and study selections. After a thorough research and examination of existing studies that leverage eBPF, cybersecurity applications around the technology were notably absent, and hence an analytical systematic review would significantly fill the existing gap of security infrastructures. The abundance of such work is constantly highlighted from the rapid expandability of eBPF in high assurance domains, including cloud native environments and microservices. This systematic literature review aims to explore the landscape of security applications that involve the eBPF technology. By providing a comprehensive analysis of the existing literature, the contributions of eBPF in the security domain can be highlighted. Moreover, it will discuss potential challenges and limitations, along with future directions in the research domain. 

The remainder of this paper is organized as follows. Section \ref{sec:2} provides background on eBPF architecture and execution semantics. Section \ref{sec:3} describes the systematic review methodology. Section \ref{results} presents the synthesized results and taxonomy of cybersecurity applications. Section \ref{patterns} discusses cross-domain patterns, challenges and research opportunities. Finally, Section \ref{concl} concludes the paper and outlines future work.

\section{Theoretical background}
\label{sec:2}
eBPF was originally designed for efficient packet filtering and has evolved into a versatile tool for system observability, networking, and security. Through just-in-time compilation and event-driven attachment to kernel hooks, eBPF supports high-performance packet processing, system tracing, and runtime observability with minimal overhead compared to user-space monitoring approaches.

A key capability relevant to cybersecurity is eBPF’s integration with high-speed packet processing paths such as the eXpress DataPath (XDP) \cite{10.1145/3281411.3281443}, which allows packet inspection and filtering to occur before traditional kernel networking stacks are traversed. This early interception enables low-latency enforcement for intrusion detection, distributed denial-of-service mitigation, and traffic policy control, while also providing fine-grained telemetry for behavioral analysis and anomaly detection across distributed environments.

Despite these advantages, eBPF programs are constrained by verifier-enforced safety guarantees, including bounded loops, restricted memory access, and limits on program complexity. While these constraints ensure kernel stability and security isolation, they also restrict the expressiveness of advanced detection logic, particularly for stateful analysis, deep packet inspection, and in-kernel ML. Consequently, many proposed cybersecurity mechanisms must balance detection capability with verifier compliance and performance scalability-an architectural tension that shapes the design space analyzed throughout this systematic review.

\section{PRISMA METHODOLOGY}
\label{sec:3}
To achieve transparency and accuracy in this systematic literature review, the PRISMA guidelines were enforced. The 2020 PRISMA flow diagram was used to guide the planning, identification and screening of the review phases of this SLR. Specifically, the diagram includes the number of records identified, retained and excluded per screening iteration and the reasons behind each decision. 

\subsection{Research Questions}
\label{sec:rqs}
\begin{figure} [h]
\centering
\includegraphics[width=0.9\linewidth]{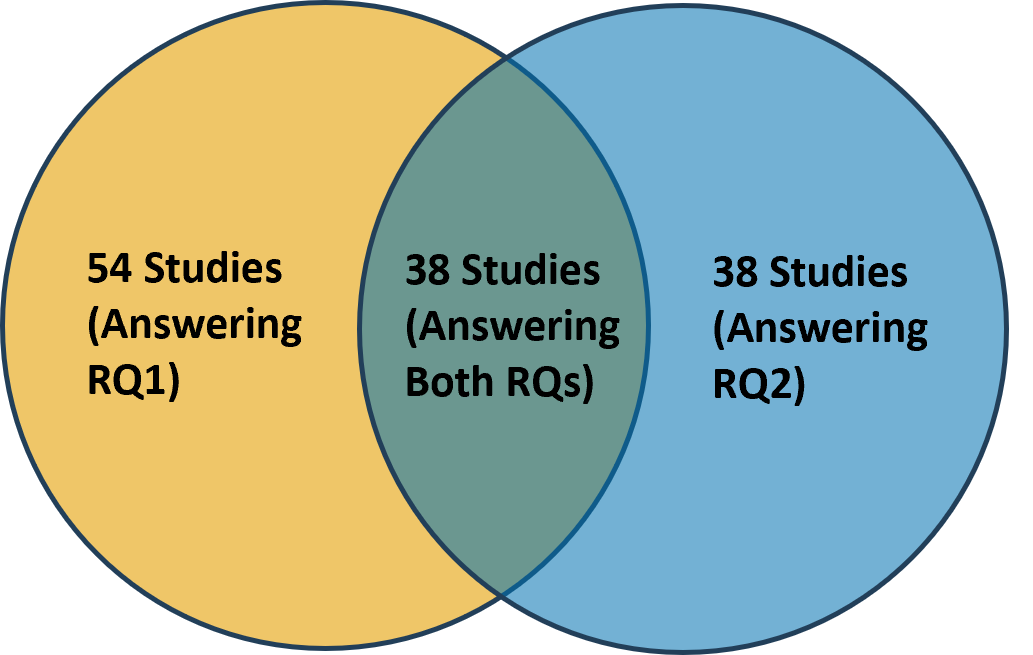}
\caption{Studies answering the RQs}
\label{RQ}
\end{figure}

To gather insights from studies, the research questions below were used for the systematic literature review. This was done to investigate the current landscape of eBPF applications regarding security. Additionally, two research questions were used to guide the inclusion and synthesis criteria of studies and further discuss the key considerations on the topic. Figure \ref{RQ} depicts the total number of selected studies that were used as intel for each of the two questions.
\begin{itemize}
    \item RQ1: How is eBPF utilized to implement cybersecurity mechanisms across different computing environments?
    \item RQ2: In what ways is eBPF applied to address specific cybersecurity threat scenarios such as DDoS, intrusion detection, and traffic analysis? 
\end{itemize}

\subsection{Eligibility Criteria}
To ensure methodological transparency and reproducibility, formal inclusion and exclusion criteria were defined prior to the screening process and applied consistently across all review phases. 

These criteria operationalize the scope boundaries implied by the research questions and are summarized in Table~\ref{tab:ie_criteria}.

\begin{table*}[ht]
\centering
\caption{Inclusion and Exclusion Criteria Applied Across Review Phases}
\label{tab:ie_criteria}
\small
\begin{tabularx}{\textwidth}{@{}llXX@{}}
\toprule
\textbf{ID} & \textbf{Criterion} & \textbf{Definition} & \textbf{Rationale} \\
\midrule
\multicolumn{4}{l}{\textit{Inclusion Criteria}} \\
\midrule
IC1 & Publication type
    & Peer-reviewed conference papers, journal articles, and books
    & Focus on validated scientific contributions; excludes informal grey literature \\
IC2 & Language
    & Publications written in English
    & Ensures consistent analysis without translation ambiguity \\
IC3 & Time frame
    & Publications from 2018 onwards
    & eBPF's modern form was introduced in kernel 4.4 (2016); 2018 marks the emergence of security-focused work \\
IC4 & Topic relevance
    & Studies that use eBPF as a primary or significant implementation mechanism for a cybersecurity objective
    & Ensures eBPF's security role is substantive, not incidental \\
IC5 & Empirical content
    & Studies presenting original empirical results, implementation details, or measurable evaluation
    & Enables extraction of performance and effectiveness data for synthesis \\
IC6 & Accessibility
    & Full text accessible via the six selected databases (Scopus, MDPI, SpringerLink, IEEE Xplore, ACM DL, ScienceDirect)
    & Ensures reproducible retrieval \\
\midrule
\multicolumn{4}{l}{\textit{Exclusion Criteria}} \\
\midrule
EC1 & Publication type
    & Review, survey, and secondary literature papers
    & Avoids circular synthesis; focuses on primary research producing novel evidence \\
EC2 & Publication date
    & Publications prior to 2018
    & Pre-2018 work predates the maturation of eBPF's security-relevant features (XDP, BTF, verifier improvements) \\
EC3 & Language
    & Non-English publications
    & Ensures consistent extraction and comparability across the corpus \\
EC4 & Peer review status
    & Theses, white papers, blog posts, preprints without peer review, and poster abstracts
    & Maintains quality threshold; these lack formal methodological validation \\
EC5 & eBPF relevance
    & Studies where eBPF appears only as a passing reference or an unrelated tool
    & Prevents dilution of findings with studies that do not meaningfully contribute eBPF security evidence \\
EC6 & Empirical quality
    & Studies lacking empirical results, insufficient technical detail, or published in low-impact venues with limited peer-review transparency
    & Applied during full-text assessment; accounts for 300 exclusions in the eligibility phase \\
EC7 & Retrievability
    & Records for which full text could not be obtained from the selected databases
    & Applied to 23 records in the retrieval phase \\
EC8 & Duplicates
    & Redundant entries across databases, identical studies with multiple index entries, and incomplete metadata records where no full text was produced
    & Removed in the identification phase ($n = 409$ duplicates; $n = 228$ other reasons) \\
\bottomrule
\end{tabularx}
\end{table*}

Criteria IC1-IC6 were applied in the order shown. IC1-IC3 governing the identification phase, IC4 and IC6 guiding title-and-abstract screening, and IC5-IC6 applied during full-text eligibility assessment. Exclusion criteria EC4 and EC8 account for the combined removal of non-peer-reviewed works ($n = 224$) and other reasons ($n = 228$) reported in the PRISMA flow diagram (Figure~\ref{fig:prisma}). Criterion EC6 corresponds to the 293 articles excluded during quality assessment in the eligibility phase. All screening decisions were finalized through a structured review process by two seperate reviewers with a third for conflicts; borderline cases were resolved through re-examination of full texts against the criteria summarized in Table~\ref{tab:ie_criteria}.

\subsection{Information Sources}
A systematic search strategy was implemented to identify relevant literature within the scope of this study. The search protocol spanned six prominent digital libraries, specifically IEEE Xplore\footnote{https://ieeexplore.ieee.org}, ACM Digital Library\footnote{https://dl.acm.org/}, Scopus\footnote{https://www.scopus.com}, SpringerLink\footnote{https://link.springer.com/}, MDPI\footnote{https://www.mdpi.com/} and Google Scholar\footnote{https://scholar.google.com/}, facilitating a robust cross-disciplinary synthesis of the field. By aggregating results from these disparate sources, the study ensures the inclusion of high-quality, peer-reviewed contributions ranging from theoretical frameworks to empirical technical evaluations.

\subsection{Search Query}

The search strategy was designed to systematically capture the full breadth of eBPF-related cybersecurity research published between 2018 and 2026. The Boolean search expression was constructed in five thematic layers, each targeting a distinct conceptual dimension of the review scope:

\begin{enumerate}
\item \textbf{Technology anchor}: "extended Berkeley Packet Filter" OR "eBPF"
\item \textbf{Security domain}: 19 terms spanning defensive mechanisms (intrusion detection, firewall, sandboxing), threat categories (DDoS mitigation, malware detection), and deployment paradigms (zero-trust, container security)
\item \textbf{Observability and performance}: Terms such as packet filtering, traffic analysis, and real-time monitoring
\item \textbf{Networking context}: Cloud-native, SDN, service mesh, and edge computing environments
\item \textbf{Application domains}: IoT, high-performance computing, and smart city infrastructure
\end{enumerate}

All five layers were connected using the Boolean AND operator, ensuring that only studies addressing eBPF within a cybersecurity context were retained. Searches were limited to publications in English, covering the period 2018-2026, and were executed independently across each of the six selected digital libraries. The complete search string is provided in Appendix~\ref{appendix:search}.

\begin{figure*}[htbp]
    \centering
    \captionsetup{justification=centering} 
    \scalebox{0.9}{ 
        \begin{tikzpicture}[>=latex, font={\sf \small}]
            \tikzstyle{bluerect} = [rectangle, rounded corners, minimum width=1.5cm, minimum height=0.75cm, text centered, draw=black, fill=cyan!60!gray!45!white, rotate=90, font=\sffamily]
            \tikzstyle{roundedrect} = [rectangle, rounded corners, minimum width=12cm, minimum height=1cm, text centered, draw=black, font=\sffamily]
            \tikzstyle{textrect} = [rectangle, minimum width=5.25cm, text width=5.24cm, minimum height=1cm, draw=black, font={\sffamily \footnotesize}]
            
            \node (top1) at (0, 10.5cm) [draw, roundedrect, fill=yellow!80!red!70]
              {\textbf{Identification of studies via database and registers}};
            
            \node (r1blue) at (-6.75cm, 8.0cm) [draw, bluerect, minimum width=3cm]{\textbf{Identification}};
            
            \node (r1left) at (-3.25cm, 8.0cm) [draw, textrect, minimum height=3cm]
              {\textbf{Records identified from Databases (n=6):}
                 \begin{itemize}
                 \item Google Scholar ($n=2810$)
                 \item IEEE Xplore ($n=211$)
                 \item ACM ($n=514$)
                 \item SpringerLink ($n=166$)
                 \item Scopus ($n=17$)
                 \item MDPI ($n=17$)
                 \newline
                 ($n=3735$)
                 \end{itemize}   
              };
            
            \node (r1right) at (3.25, 8.0cm) [draw, textrect, minimum height=3cm]
              {\textbf{Records removed before screening:} 
                \begin{itemize}
                \item Duplicates ($n=409$)
                \item Published before 2018 ($n=127$)
                \item Non English ($n=66$)
                \item Non peer-reviewed ($n=259$)
                \item Other reasons ($n=228$)
                \newline
                ($n=1089$)
                \end{itemize} 
              };
            
            \node (r2blue) at (-6.75cm, 1.5cm) [draw, bluerect, minimum width=8.5cm]
              {\textbf{Screening}};
            
            \node (r2left) at (-3.25cm, 5.0cm) [draw, textrect, minimum height=1.5cm]
              {\textbf{Records screened ($n=2646$)}};
            
            \node (r2right) at (3.25, 5.0cm) [draw, textrect, minimum height=1.5cm]
              {\textbf{Records excluded based on title, abstract ($n=2267$)}};
            
            \node (r3left) at (-3.25cm, 2.5cm) [draw, textrect, minimum height=1.5cm]
              {\textbf{Reports sought for retrieval ($n=379$)}};
            
            \node (r3right) at (3.25, 2.5cm) [draw, textrect, minimum height=1.5cm]
              {\textbf{Not retrieved ($n=23$)}};

            \node (r4left) at (-3.25cm, 0cm) [draw, textrect, minimum height=1.5cm]
              {\textbf{Reports assessed for eligibility \\ ($n=356$)}};
            
            \node (r4right) at (3.25, -1.0cm) [draw, textrect, minimum height=3.5cm]
              {\textbf{Reports Excluded:} 
                \begin{itemize}
                \item Review Papers ($n=5$)
                \item Excluded based on quality assessment: lack of empirical results or insufficient technical detail ($n=297$)
                \end{itemize}     
              };

            \node (r5blue) at (-6.75cm, -4.5cm) [draw, bluerect, minimum width=2cm]
              {\textbf{Included}};
            
            \node (r5left) at (-3.25cm, -4.5cm) [draw, textrect, minimum height=2cm]
              {\textbf{Studies included in review} \\ 
               $n=54$ \\ 
               };
            
            \draw[thick, ->] (r1left) -- (r1right);
            \draw[thick, ->] (r1left) -- (r2left);
            \draw[thick, ->] (r2left) -- (r2right);
            \draw[thick, ->] (r2left) -- (r3left);
            \draw[thick, ->] (r3left) -- (r3right);
            \draw[thick, ->] (r3left) -- (r4left);
            \draw[thick, ->] (r4left) -+ (0.5,0);
            \draw[thick, ->] (r4left) -- (r5left);
        \end{tikzpicture}
    }
    \caption{Overview of the literature research procedure based on PRISMA principles}
    \label{fig:prisma}
\end{figure*}
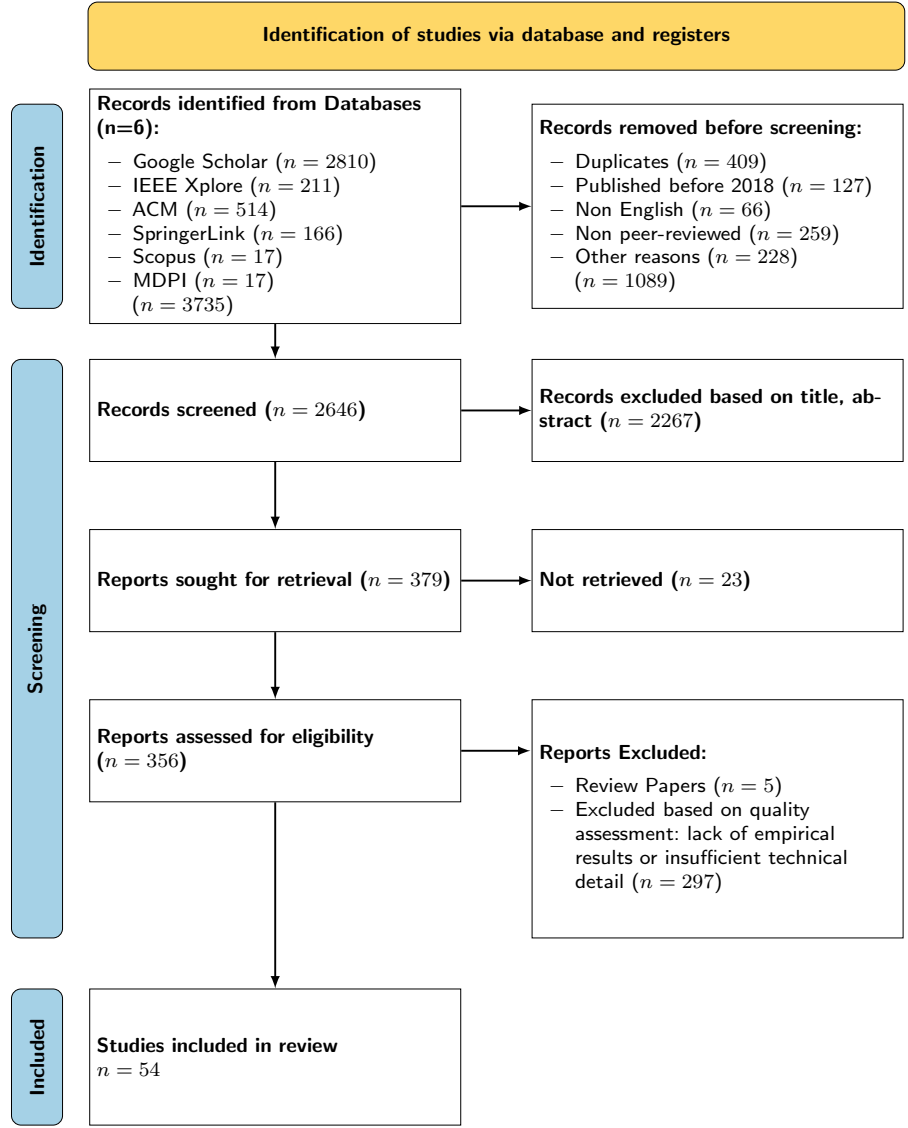

\subsection{Article Selection}

The initial identification phase yielded a total of 3735 records from six primary databases. To ensure the relevance and technical applicability of the findings to modern systems, we excluded 127 titles published before 2018. Further refinement during this stage involved removing duplicates (n=409), non-English publications (n=66), and non-peer-reviewed works such as certain theses or white papers (n=259), other reasons category (n=228) primarily accounted for redundant entries across different database versions, incomplete metadata entries where full-texts were never produced, and short-form poster abstracts that did not contain sufficient technical depth for data. extraction resulting in 2646 records for initial screening. During the title and abstract screening, 2267 sources were excluded for being outside the cybersecurity domain, leaving 379 reports sought for retrieval. Of these, 23 could not be fully retrieved or lacked sufficient access, resulting in 356 reports for full-text eligibility assessment. In this final stage, we excluded 5 review papers to focus on primary research and removed 297 studies based on a rigorous quality assessment. These exclusions were primarily due to a lack of empirical results, insufficient technical detail, or publication in low-impact venues with limited peer-review transparency. This process concluded with 54 high-quality data sources being selected as the primary literature for this review.

This multi-stage attrition process ensured that only studies exhibiting high methodological transparency and verifiable experimental frameworks were retained. By applying these stringent exclusion criteria, the final corpus was distilled to 54 high-quality primary sources that constitute the foundational evidence for this review. 

\subsection{Analysis of Review Phases}
Since eBPF is a widely used technology, it appears as the primary framework across a diverse range of system-level implementations, including performance monitoring, load balancing, packet processing, and dynamic tracing. Despite this, its application to cybersecurity remains a distinct and underexplored dimension of the literature. The aim of this review is therefore to focus specifically on security-oriented eBPF applications operating at the kernel level, drawing from peer-reviewed contributions accessible through licensed and established research databases. Three screening phases were conducted across the selected databases, as illustrated in Figure~\ref{fig:prisma}, to maximise coverage and minimise selection bias.

\begin{figure}[h]
\centering
\includegraphics[width=0.9\linewidth]{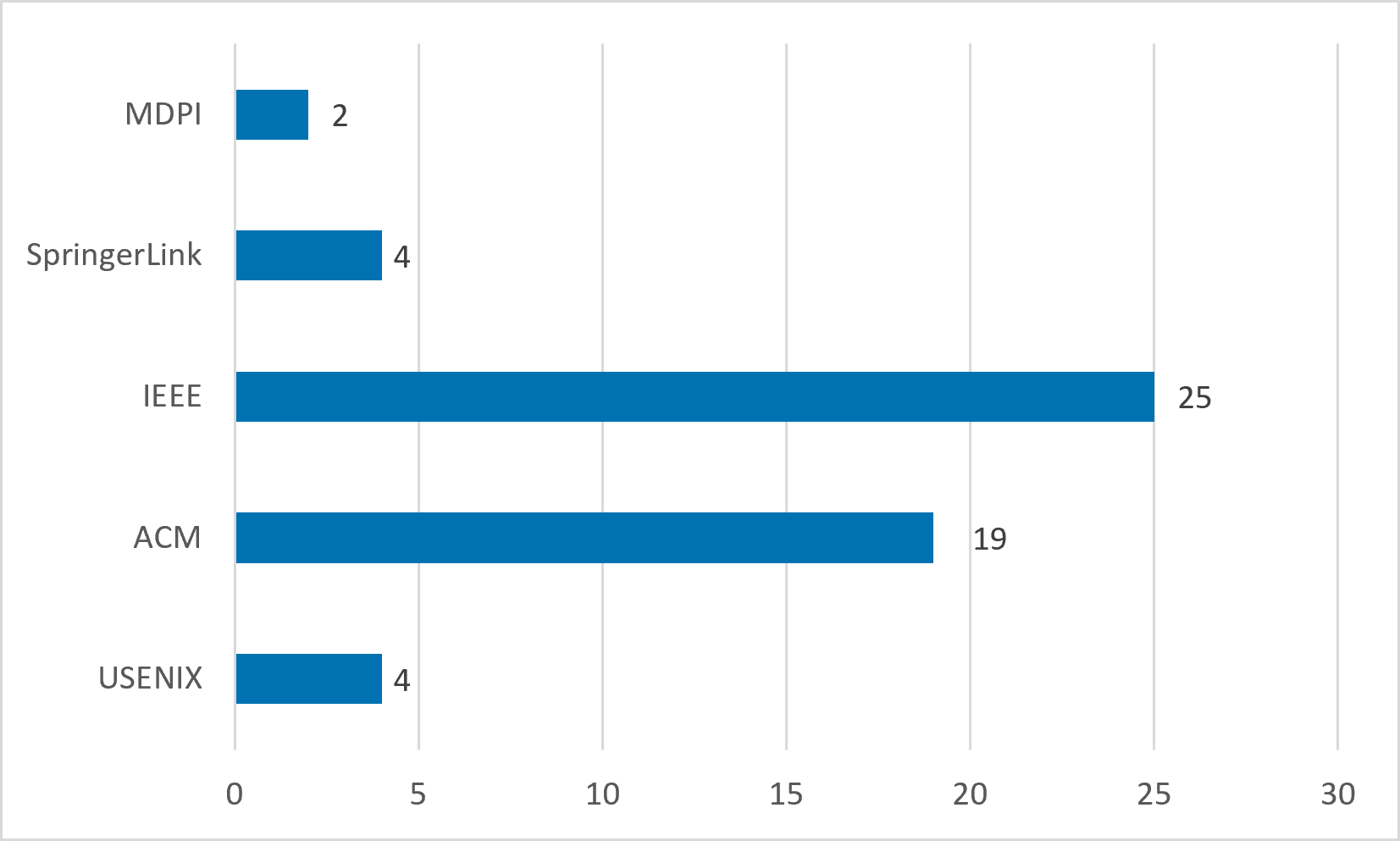}
\caption{Distribution among databases}
\label{data}
\end{figure}

The corpus exhibits strong concentration as shown in Figure \ref{data} in top-tier systems and security venues, with IEEE Xplore (25 papers, 46.2\%) and ACM Digital Library (19 papers, 35.1\%) together accounting for 81.3\% [44/54] of all studies. The remaining publications were distributed across SpringerLink (7.4\%) [4/54], USENIX (7.4\%) [4/54], and MDPI open-access journals and conferences (3.7\%) [2/54]. This distribution reflects eBPF security research's position within the systems and networking community.

\begin{figure} [h]
\centering
\includegraphics[width=0.9\linewidth]{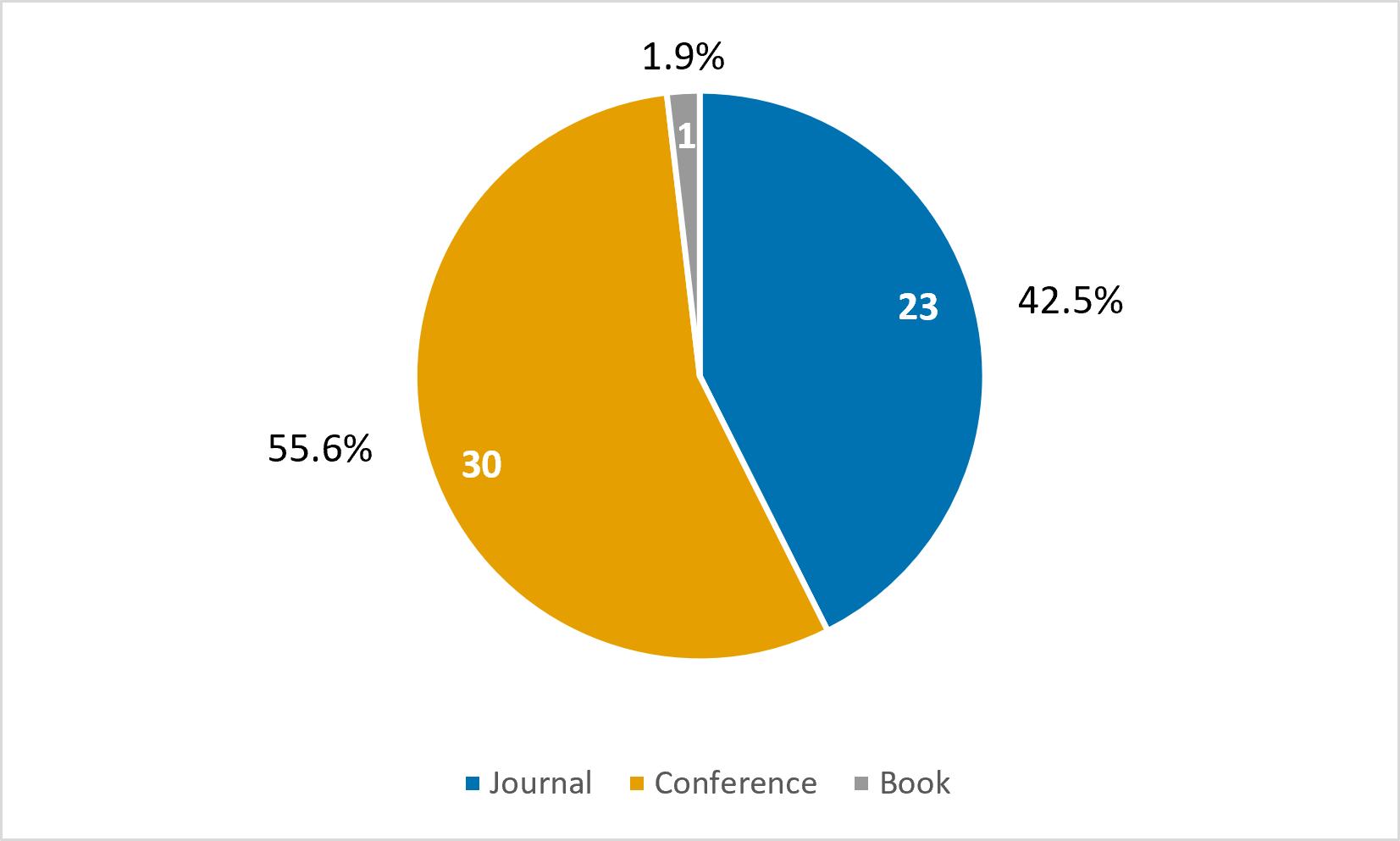}
\caption{Distributions among conferences, journals, books}
\label{conf}
\end{figure}

As shown in Figure~\ref{conf}, a total of 54 studies successfully passed all screening steps and constitute the primary evidence base for this SLR. The corpus is distributed across three publication types: (i) 30 conference papers on cybersecurity topics (55\%); (ii) 23 journal articles (43\%); 
and (iii) 1 book chapter (2\%). The dominance of conference papers reflects the fast-moving nature of eBPF research, where novel contributions are typically disseminated rapidly through peer-reviewed venues before maturing into journal publications.

\begin{figure} [h]
\centering
\includegraphics[width=0.9\linewidth]{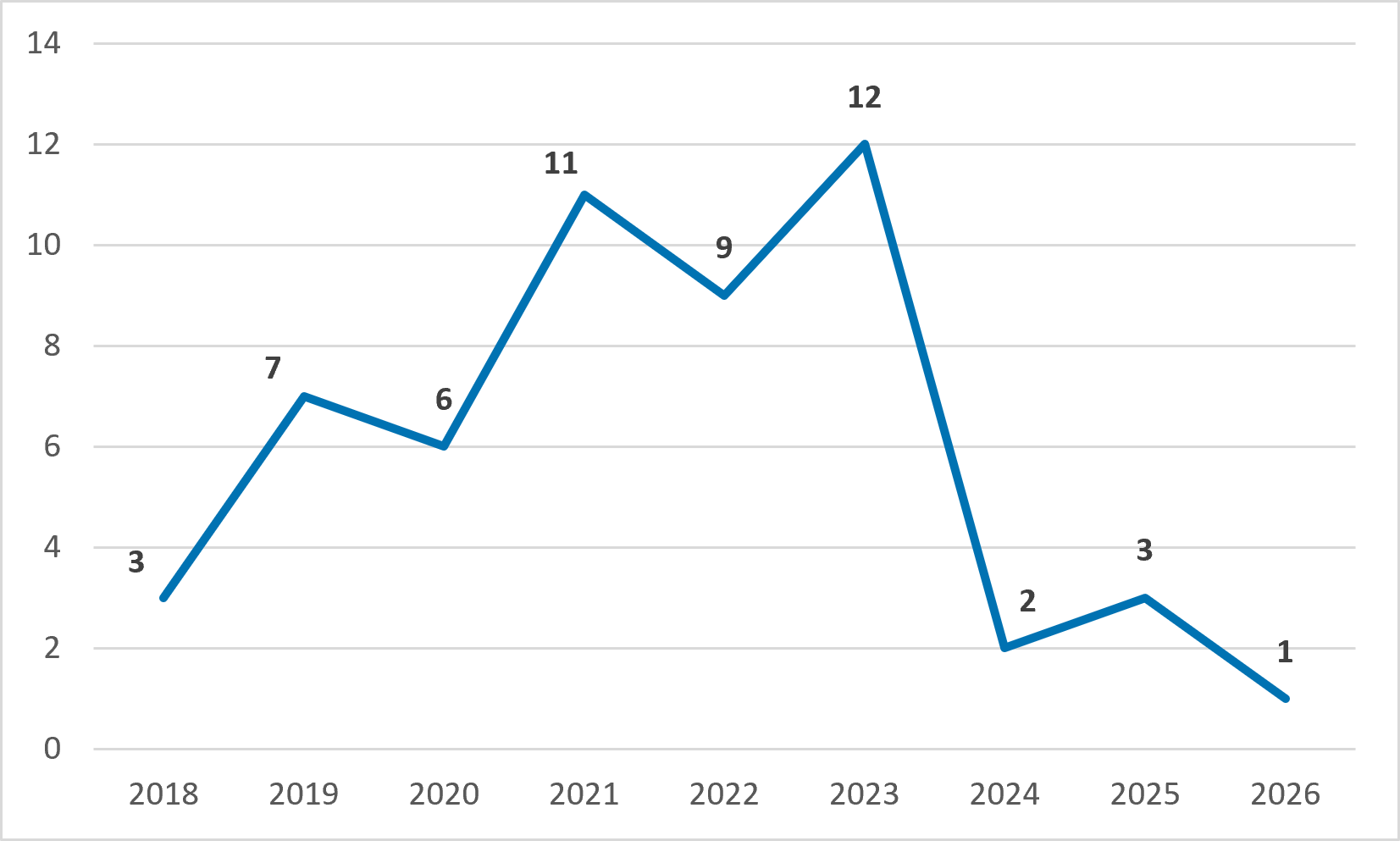}
\caption{Publishing years}
\label{year}
\end{figure}

Figure~\ref{year} illustrates the temporal distribution of publications within the scope of this SLR. Output was modest in the early period, with 3 publications in 2018 and 7 in 2019, before accelerating sharply to 11 in 2021 along with 12 in 2023. This trajectory closely mirrors the stabilisation of key eBPF capabilities in the Linux kernel including XDP, BTF, and bounded loop support suggesting that research activity is responsive to platform maturation. The comparatively lower counts for 2024-2026 are attributable to the timeframe of data collection and the standard indexing lag inherent to academic databases, rather than a contraction of research interest. Taken together, the temporal distribution confirms the timeliness and relevance of this systematic review.

\section{Results and Discussion}
\label{results}
To systematically synthesize the reviewed literature, a thematic coding process was applied to the extracted study attributes, including security objective, deployment layer, enforcement mechanism, and evaluation context. Each of the 54 primary studies was assigned to a functional domain reflecting its dominant eBPF usage, and the resulting taxonomy was validated against the two research questions defined in Section~\ref{sec:rqs}.

RQ1, examining how eBPF is utilized to implement cybersecurity mechanisms across different computing environments, is addressed across all seven thematic categories. Each category represents a distinct deployment context: from resource-constrained IoT edge devices and containerised cloud workloads, 
to distributed microservice architectures, high-speed networking stacks, and general-purpose security tooling. The breadth of environments covered demonstrates eBPF's architectural versatility as a kernel-level programmability layer that transcends any single deployment paradigm.

RQ2, examining how eBPF addresses specific cybersecurity threat scenarios such as DDoS, intrusion detection, and traffic analysis, is addressed primarily in Sections~\ref{sec:ddos} (DDoS Mitigation), \ref{sec:ids} (Intrusion Detection), \ref{sec:containers} (Container Security), and \ref{sec:microservices} (Microservice Protection), where the reviewed works document concrete threat models, enforcement mechanisms, and empirical evaluations against named attack classes. Partial contributions to RQ2 also emerge from the Networking and Tools \& Frameworks categories, particularly 
in studies addressing traffic privacy, covert channel detection, and vulnerability profiling.

Together, the 54 studies entries cataloged in Tables~\ref{tab:primary_studies_1} and \ref{tab:primary_studies_2} span seven domains and cover publications from 2018 to 2026, reflecting both the maturation of eBPF's security-relevant capabilities and the growing diversity of threat scenarios to which it has been applied. The distribution of studies across domains, with IDS representing the largest category ($n=11$, 20.3\%), followed by Container Security ($n=10$, 18.5\%)and Microservices ($n=6$ each, 11.1\%), DDoS ($n=7$, 12.9\%), Tools \& Frameworks ($n=9$, 16.6\%), IoT ($n=6$, 11.1\%), and Networking ($n=5$,9.2\%), reflects the research community's prioritisation of detection and cloud-native security over protocol-level and infrastructure concerns. This analysis revealed seven recurring functional domains that collectively define the design space of eBPF-enabled cybersecurity, each discussed in the 
subsections that follow.

Because several studies overlap multiple domains, each paper was assigned to a single category based on its primary research objective rather than its evaluation environment. For instance, DDoS-specific detection and mitigation studies were classified as DDoS, whereas general attack detection mechanisms were classified as IDS. Likewise, works targeting container runtime or orchestration security were categorized as Container Security, while those addressing service-level security, communication, or policy enforcement in microservice architectures were categorized as Microservices.
Each paper fell into its most relevant category based on eBPF usage and its dominant security role. Studies that combine more than 1 categories are discussed in Section ~\ref{sec:5}.

It is also essential to clarify that, even though papers were categorized in the aforementioned 7 categories, security telemetry is treated as a cross-cutting capability between multiple domains rather than a standalone application domain. Solutions do use eBPF to collect runtime data, such as kernel events, syscall traces and network flow information, which are all subsequently used for detection, observability and policy enforcement.
\begin{table*}[!h]
\centering
\caption{Primary Studies Included in the Systematic Literature Review (Part 1 of 2)}
\label{tab:primary_studies_1}
\small
\begin{tabularx}{\textwidth}{@{} p{0.8cm} p{0.9cm} X p{2.6cm} p{0.8cm} p{1.6cm} @{}}
\toprule
\textbf{No.} & \textbf{Year} & \textbf{Paper Title} & \textbf{Domain} & \textbf{Ref.} & \textbf{RQ} \\
\midrule

1  & 2022 & Signature-Based Detection of Botnet DDoS Attacks
           & DDoS & \cite{Szynkiewicz2022}  & RQ1, RQ2 \\[4pt]
2  & 2021 & SYN Flood Attack Detection and Mitigation Using Machine Learning Traffic Classification and Programmable Data Plane Filtering
           & DDoS & \cite{Dimolianis2021126}  & RQ1, RQ2 \\[4pt]
3  & 2019 & Introducing SmartNICs in Server-Based Data Plane Processing: The DDoS Mitigation Use Case
           & DDoS & \cite{0c6699b190594dfd8bcd14b8828899b5}  & RQ1, RQ2 \\[4pt]
4  & 2020 & A Multi-Feature DDoS Detection Schema on P4 Network Hardware
           & DDoS & \cite{Dimolianis20201} & RQ1, RQ2 \\[4pt]
5  & 2023 & Coda: Runtime Detection of Application-Layer CPU-Exhaustion DoS Attacks in Containers
           & DDoS & \cite{9842371} & RQ1, RQ2 \\[4pt]
6  & 2019 & Detecting Asymmetric Application-Layer Denial-of-Service Attacks In-Flight with FineLame
           & DDoS & \cite{Demoulin2019693} & RQ1, RQ2 \\[4pt]
7  & 2021 & DDoS Attack and Defense in SDN-Based Cloud
           & DDoS & \cite{10.1007/978-3-030-86356-2_13} & RQ1, RQ2 \\
\midrule

8  & 2022 & Design and Implementation of an Intrusion Detection System by Using Extended BPF in the Linux Kernel
           & IDS & \cite{WANG2022103283} & RQ1, RQ2 \\[4pt]
9  & 2019 & LBM: A Security Framework for Peripherals within the Linux Kernel
           & IDS & \cite{Tian2019967} & RQ1, RQ2 \\[4pt]
10 & 2021 & An End-to-End Framework for Machine Learning-Based Network Intrusion Detection System
           & IDS & \cite{9501960} & RQ1, RQ2 \\[4pt]
11 & 2022 & Enhancing Port Scans Attack Detection Using Principal Component Analysis and Machine Learning Algorithms
           & IDS & \cite{10.1007/978-981-19-8445-7_8} & RQ1, RQ2 \\[4pt]
12 & 2022 & Real-Time Heuristic-Based Detection of Attacks Performed on a Linux Machine Using Osquery
           & IDS & \cite{Ahamed2022RealTimeHD} & RQ1, RQ2 \\[4pt]
13 & 2020 & Spectre Attacks: Exploiting Speculative Execution
           & IDS & \cite{10.1145/3399742} & RQ1 \\[4pt]
14 & 2022 & A Scalable and Dynamic ACL System for In-Network Defense
           & IDS & \cite{Jung20221679} & RQ1, RQ2 \\[4pt]
15 & 2021 & An Autonomous Cybersecurity Framework for Next-Generation Digital Service Chains
           & IDS & \cite{article2} & RQ1, RQ2 \\[4pt]
16 & 2023 & Guarding Serverless Applications with Kalium
           & IDS & \cite{Jegan20234087} & RQ1, RQ2 \\[4pt]
17 & 2023 & PALANTIR: An NFV-Based Security-as-a-Service Approach for Automating Threat Mitigation
           & IDS & \cite{s23031658} & RQ1, RQ2 \\[4pt]
18 & 2021 & eZTrust: Network-Independent Zero-Trust Perimeterization for Microservices
           & IDS & \cite{10.1145/3314148.3314349} & RQ1, RQ2 \\
\midrule

19 & 2019 & Secure Edge Computing with Lightweight Control-Flow Property-Based Attestation
           & IoT & \cite{8806658} & RQ1 \\[4pt]
20 & 2019 & Multi-Layer IoT Security Framework for Ambient Intelligence Environments
           & IoT & \cite{s19184038} & RQ1 \\[4pt]
21 & 2018 & Towards In-Network Security for Smart Homes
           & IoT & \cite{10.1145/3230833.3232802} & RQ1 \\[4pt]
22 & 2023 & Design and Implementation of a Sandbox for Facilitating and Automating IoT Malware Analysis with Techniques to Elicit Malicious Behavior
           & IoT & \cite{article1} & RQ1, RQ2 \\[4pt]
23 & 2020 & Secure Software-Defined Networking Communication Systems for Smart Cities: Current Status, Challenges, and Trends
           & IoT & \cite{Rahouti202112083} & RQ1 \\[4pt]
24 & 2021 & Integrating Ultra-Wideband and Free Space Optical Communication for Realizing a Secure and High Throughput Body Area Network Architecture
           & IoT & \cite{unknown} & RQ1 \\
\midrule

25 & 2021 & Secure Namespaced Kernel Audit for Containers
           & Container Sec. & \cite{10.1145/3472883.3486976} & RQ1, RQ2 \\[4pt]
26 & 2020 & The Rise of eBPF for Non-Intrusive Performance Monitoring
           & Container Sec. & \cite{9110434} & RQ1 \\[4pt]
27 & 2020 & Rule-Based Security Monitoring of Containerized Environments
           & Container Sec. & \cite{10.1007/978-3-030-49432-2_4} & RQ1, RQ2 \\
28 & 2021 & Container Hardening Through Automated Seccomp Profiling
           & Container Sec. & \cite{10.1145/3429885.3429966} & RQ1, RQ2 \\[4pt]
29 & 2018 & A Measurement Study on Linux Container Security: Attacks and Countermeasures
           & Container Sec. & \cite{10.1145/3274694.3274720} & RQ1, RQ2 \\[4pt]
30 & 2025 & Detecting Cryptojacking Containers Using eBPF-Based Security Runtime and Machine Learning
           & Container Sec. & \cite{electronics14061208} & RQ1, RQ2 \\[4pt]
31 & 2026 & Hybrid Runtime Detection of Malicious Containers Using eBPF
           & Container Sec. & \cite{RYU2026} & RQ1, RQ2 \\[4pt]
\bottomrule
\end{tabularx}-
\end{table*}

\begin{table*}[t]
\centering
\caption{Primary Studies Included in the Systematic Literature Review (Part 2 of 2)$^{*}$}
\label{tab:primary_studies_2}
\small
\begin{tabularx}{\textwidth}{@{} p{0.5cm} p{0.7cm} X p{2.4cm} p{0.6cm} p{1.4cm} @{}}
\toprule
\textbf{No.} & \textbf{Year} & \textbf{Paper Title} & \textbf{Domain} & \textbf{Ref.} & \textbf{RQ} \\
\midrule

32 & 2023 & Cross Container Attacks: The Bewildered eBPF on Clouds
           & Container Sec. & \cite{ADDED_PAPER3} & RQ1, RQ2 \\
33 & 2024 & P4Control: Line-Rate Cross-Host Attack Prevention via In-Network Information Flow Control Enabled by Programmable Switches and eBPF
           & Container Sec. & \cite{ADDED_PAPER2} & RQ1, RQ2 \\
34 & 2022 & Secure Inter-Container Communications Using XDP/eBPF
            & Container Sec. & \cite{ADDED_PAPER1} & RQ1, RQ2 \\
\midrule
35 & 2022 & Learning State Machines to Monitor and Detect Anomalies on a Kubernetes Cluster
           & Microservices & \cite{10.1145/3538969.3543810} & RQ1, RQ2 \\[4pt]
36 & 2023 & Microservice Security Metrics for Secure Communication, Identity Management, and Observability
           & Microservices & \cite{Zdun2023} & RQ1, RQ2 \\[4pt]
37 & 2023 & $\mu$Detector: Automated Intrusion Detection for Microservices
           & Microservices & \cite{Flora2023748} & RQ1, RQ2 \\[4pt]
38 & 2021 & A Framework for eBPF-Based Network Functions in an Era of Microservices
           & Microservices & \cite{9340283} & RQ1, RQ2 \\[4pt]
39 & 2021 & Kmon: An In-Kernel Transparent Monitoring System for Microservice Systems with eBPF
           & Microservices & \cite{v} & RQ1, RQ2 \\[4pt]
40 & 2025 & AIDS-Based Cyber Threat Detection Framework for Secure Cloud-Native Microservices
           & Microservices & \cite{electronics14020229} & RQ1, RQ2 \\
\midrule

41 & 2022 & Leveraging eBPF to Make TCP Path-Aware
           & Networking & \cite{9772044} & RQ1 \\[4pt]
42 & 2024 & TANGO: Secure Collaborative Route Control Across the Public Internet
           & Networking & \cite{Birge-Lee20241791} & RQ1 \\[4pt]
43 & 2023 & MiddleNet: A Unified, High-Performance NFV and Middlebox Framework With eBPF and DPDK
           & Networking & \cite{10068810} & RQ1 \\[4pt]
44 & 2018 & Performance Implications of Packet Filtering with Linux eBPF
           & Networking & \cite{inproceedings} & RQ1 \\[4pt]
45 & 2020 & Leveraging eBPF to Preserve User Privacy for DNS, DoT, and DoH Queries
           & Networking & \cite{10.1145/3407023.3407041} & RQ1, RQ2 \\
\midrule

46 & 2021 & Kernel-Level Tracing for Detecting Stegomalware and Covert Channels in Linux Environments
           & Tools \& Fr. & \cite{CAVIGLIONE2021108010} & RQ1, RQ2 \\[4pt]
47 & 2023 & SysFlow: Toward a Programmable Zero Trust Framework for System Security
           & Tools \& Fr. & \cite{10091151} & RQ1, RQ2 \\[4pt]
48 & 2023 & Full-Stack Vulnerability Analysis of the Cloud-Native Platform
           & Tools \& Fr. & \cite{ZENG2023103173} & RQ1 \\[4pt]
49 & 2019 & Simple and Precise Static Analysis of Untrusted Linux Kernel Extensions
           & Tools \& Fr. & \cite{10.1145/3314221.3314590} & RQ1 \\[4pt]
50 & 2023 & vkTracer: Vulnerable Kernel Code Tracing to Generate Profile of Kernel Vulnerability
           & Tools \& Fr. & \cite{10.1007/978-3-031-25659-2_16} & RQ1, RQ2 \\[4pt]
51 & 2021 & Synthesizing Safe and Efficient Kernel Extensions for Packet Processing
           & Tools \& Fr. & \cite{Xu202150} & RQ1 \\[4pt]
52 & 2023 & Seeing the Invisible: Auditing eBPF Programs in Hypervisor with HyperBee
           & Tools \& Fr. & \cite{10.1145/3609021.3609305} & RQ1, RQ2 \\[4pt]
53 & 2022 & A Control Plane Enabling Automated and Fully Adaptive Network Traffic Monitoring With eBPF
           & Tools \& Fr. & \cite{9869628} & RQ1 \\[4pt]
54 & 2025 & Aeolia: A Fast and Secure Userspace Interrupt-Based Storage Stack
           & Tools \& Fr. & \cite{li2025aeolia} & RQ1 \\

\bottomrule
\multicolumn{6}{l}{}%
\end{tabularx}
\end{table*}

\subsection{Distributed Denial of Service detection and mitigation}
\label{sec:ddos}
Modern DDoS attacks combine volumetric flooding, application-layer resource exhaustion, and polymorphic signatures that evade static defenses. eBPF's ability to process packets at earliest stages through XDP and monitor system resources enables sophisticated multi-tier defense strategies.

\subsubsection{Signature-Based Filtering}
Traditional signature-based defenses remain relevant when enhanced with in-kernel programmability. Researchers \cite{Szynkiewicz2022} highlight the sophistication of modern DDoS attacks and propose a novel signature-based mitigation technique leveraging Packet Generation Algorithms (PGAs). The method uses eBPF/XDP for fast, protocol-level detection of malicious packets, offering supplementary protection against evolving DDoS threats. Early packet filtering at the XDP hook enables line-rate processing before kernel networking stack allocation.

Following the flood identification approach, authors \cite{Dimolianis2021126} propose a detection and mitigation schema focusing on TCP SYN flood attacks. Machine learning models classify network traffic into malicious and benign signatures, which are then enforced by high-end eBPF/XDP firewalls. The system optimizes signatures through a reduction process with two objectives: minimizing malicious signature count and reducing collateral damage to legitimate traffic. Results demonstrate high detection accuracy with better performance than SYN cookies mechanisms in high-speed scenarios.

\subsubsection{Hardware-Software Hybrids}
To combat high-speed attacks, recent approaches split processing between hardware acceleration and kernel-level agility. Study \cite{0c6699b190594dfd8bcd14b8828899b5} explores integration of eBPF/XDP with SmartNICs to enhance network data plane efficiency. The approach offloads portions of DDoS mitigation rules to SmartNICs for hardware-accelerated filtering, while using XDP in the kernel for flexible traffic sampling and aggregation. Performance evaluations demonstrate that combining SmartNIC offloading with host kernel processing optimizes packet handling for efficient DDoS mitigation.

Complementing this work, researchers in \cite{Dimolianis20201} address gaps in system evaluations with recorded traffic data. Their in-network architecture combines traffic metrics (flow counts, packet symmetry) evaluated in SmartNICs to identify anomalies. When anomalies are detected, eBPF scripts trigger alerts and pass traffic to external mitigation systems. The approach demonstrates high packet processing rates (1-2 Mbps for 100G links) with strong detection accuracy using P4-enabled SmartNICs.

\subsubsection{Application-Layer Defenses}
Beyond network-layer attacks, stealthy CPU-exhaustion attacks demand deeper runtime visibility. Researchers in \cite{9842371} present Coda, a framework for detecting application-layer CPU-exhaustion DoS attacks in containers. Unlike traditional volumetric attacks, CPU-exhaustion exploits application vulnerabilities that are difficult to detect with network-layer defenses. Coda monitors per-connection CPU time consumption using eBPF to trace system calls at the host level, applying statistical methods to detect anomalous patterns. The framework is language-agnostic, transparent to containerized applications, and demonstrates accurate detection with minimal overhead.

Study \cite{Demoulin2019693} introduces FILENAME, addressing asymmetric DoS attacks where request processing costs significantly exceed attacker effort. The framework leverages OS-level visibility and eBPF to attach user-level probes to request processing functions and lightweight resource monitors to allocation functions. This data trains models of normal resource utilization, enabling detection of resource abuse attacks with low overhead across legacy applications.

\subsubsection{SDN-Based Mitigation}
In cloud environments, distributed programmability enables scalable defenses. Research in \cite{10.1007/978-3-030-86356-2_13} proposes an efficient architecture based on OpenStack, ONOS controller, and Open vSwitch. A mitigation mechanism incorporates uBPF (user-space eBPF implementation) into reconfigurable data paths, offering runtime program execution. The system leverages switch programmability, distributed packet processing, and centralized SDN control to resist DDoS flood attacks in cloud architectures.

The reviewed DDoS studies reveal a progression from simple packet filtering toward sophisticated multi-tier defenses combining hardware acceleration, kernel-level monitoring, and application-aware resource tracking, all enabled by eBPF's flexible deployment model.

\subsection{Intrusion detection systems}
\label{sec:ids}
Traditional signature-based IDS suffer from high false positive rates and inability to detect zero-day attacks. eBPF enables a new generation of IDS that combine machine learning, behavioral analysis, and hardware integration for sophisticated threat detection.

Early eBPF-based IDS research established foundational architectures combining kernel-space and user-space processing. Authors \cite{WANG2022103283} present an IDS leveraging eBPF for efficient packet inspection directly in the Linux kernel. The system employs a split architecture: kernel-space eBPF module for rapid pattern matching to filter irrelevant packets, and user-space module for in-depth rule-based analysis of remaining packets. Experimental evaluations using modified Snort rulesets demonstrate up to 3× throughput improvement versus traditional Snort under various traffic conditions, validating eBPF's performance advantages for network-based intrusion detection.

Peripheral Security Framework. Expanding IDS scope beyond network traffic, study \cite{Tian2019967} introduces Linux (e)BPF Modules (LBM), a security framework providing unified API for protecting against malicious peripherals within the Linux kernel. LBM leverages eBPF's packet filtering functionality for performance and extensibility, offering a high-level language for developing filtering mechanisms. The framework demonstrates host protection against malicious USB, Bluetooth, and NFC devices with overhead approaching 1$\mu$s per packet, establishing LBM as the first kernel-based security framework addressing peripheral threats comprehensively.

Study \cite{9501960} introduces the AB-TRAP framework, designed for ML-based network IDS using updated traffic datasets. The framework includes dataset generation, model training, implementation, and post-deployment evaluation. Applied to TCP port scan detection in LAN and Internet environments, AB-TRAP utilizes eBPF for high-throughput packet capture and real-time model updates. Results demonstrate high F1-scores with minimal resource usage, addressing deployment concerns through reproducible methodologies.

Research \cite{10.1007/978-981-19-8445-7_8} deploys seven ML classifiers with principal component analysis for port scan detection. The study compares performance metrics (accuracy, precision, recall, AUC, F1-score, false positive rate) across models, identifying XGBoost as the best classifier with 99.98\% accuracy. eBPF provides efficient packet monitoring and feature extraction for ML model training and inference.

Authors \cite{Ahamed2022RealTimeHD} focus on detecting attacks at the Tactics, Techniques, and Procedures (TTPs) level of David Bianco's Pyramid of Pain. Using heuristic-based ML models with osquery integration, the system detects anomalies missed by traditional signatures. eBPF enables fast pattern matching while pre-dropping packets that don't match initial rule sets, facilitating detection of lateral movement and privilege escalation.

Study \cite{10.1145/3399742} addresses Spectre attacks exploiting modern CPU speculative execution. These attacks can read arbitrary memory and bypass process separation, containerization, and JIT compilation protections. eBPF technology enables unprivileged users to access JIT compilation modules where Spectre vulnerabilities are exposed, demonstrating the need for microarchitectural-level monitoring.

Research \cite{Jung20221679} proposes PortCatcher, an in-switch Access Control List (ACL) system supporting autonomous defenses at scale. The system introduces linear range maps (LRM) enabling port matching in SRAM-based hash tables, combined with eBPF for low-latency ACL management. PortCatcher achieves significant TCAM space savings (74-90\%) while maintaining rapid rule deployment for attack blocking.

Study \cite{article2} proposes a comprehensive framework for heterogeneous security services across multi-domain distributed systems. Key innovations include full automation of threat detection and response processes, dynamic adaptation to emerging attack patterns, and real-time adjustment of monitoring granularity using eBPF. The framework addresses mutual trustworthiness, advanced threat mitigation, identity management, and access control across cloud, edge, and on-premises environments.

Research \cite{Jegan20234087} introduces Kalium, an extensible security framework leveraging local and global application states to enhance control-flow integrity in serverless applications. Utilizing eBPF for performance optimization, Kalium constantly monitors system resource utilization and demonstrates attack mitigation with low overhead, outperforming existing serverless information flow protection systems.

The PALANTIR project \cite{s23031658} addresses cybersecurity challenges for small and medium enterprises through outsourced security supervision. The architecture extends NFV to automate threat mitigation and remediation using eBPF's in-kernel networking functionalities. Key contributions include ontology-based decision-making, remediation policy models, and adaptable deployment across diverse environments.

IDS research demonstrates eBPF's evolution from simple packet filtering toward sophisticated ML-powered, behaviorally-aware, and hardware-integrated threat detection systems.

\subsection{Internet of Things Security}
The exponential growth of IoT devices has outpaced traditional security models, forcing researchers to rethink protection strategies for resource-constrained, heterogeneous environments. eBPF's lightweight, programmable nature makes it particularly suitable for IoT security, where computational overhead and power consumption are critical constraints.

\subsubsection{Decentralized Attestation}
Researchers \cite{8806658} address critical IoT security challenges by proposing a lightweight dynamic control-flow property-based attestation (CFPA) architecture. Designed for resource-constrained edge devices and cloud systems, CFPA verifies software integrity by attesting to critical, small, and simple software components. This decentralized approach eliminates the need for federated servers or trusted third parties, leveraging eBPF to monitor control-flow execution with minimal overhead. The framework enhances efficiency and scalability for edge computing scenarios where centralized trust models prove impractical.

\subsubsection{Healthcare IoT Frameworks}
The healthcare IoT domain presents unique security requirements due to sensitive patient data and real-time monitoring needs. Study \cite{s19184038} introduces a modular three-layer architecture integrating eBPF for secure health data management. The framework separates device management, data aggregation, and security policy enforcement, enabling flexible deployment across diverse IoT ecosystems. eBPF programs enforce access control policies at network boundaries while preserving end-to-end encryption for medical data transmission.

Building on healthcare IoT security, researchers in \cite{10.1145/3230833.3232802} propose an enhanced framework incorporating additional security layers for ambient intelligence environments. The multi-layer approach addresses challenges in smart home and assisted living scenarios, where heterogeneous devices must interoperate securely. eBPF facilitates dynamic security policy updates without device firmware modifications, critical for long-lived IoT deployments.

Study \cite{article1} explores integration of ultra-wideband (UWB) body area networks with free-space optical (FSO) communication channels for secure health data transmission. eBPF programs filter and inspect network traffic at UWB-BAN interfaces, ensuring secure transactions between health monitoring devices and centralized systems. The architecture demonstrates low bit error rates (BER) with cost-efficient deployment, leveraging eBPF for real-time traffic inspection without compromising communication latency.

\subsubsection{Smart City Security}
Extending eBPF's applicability to urban-scale deployments, research \cite{Rahouti202112083} demonstrates Software-Defined Networking (SDN) integration for smart city security. The work addresses security challenges in smart city applications including transportation and healthcare, proposing SDN-based solutions leveraging eBPF for programmable security enforcement across distributed city infrastructure. The survey identifies future research trends in smart buildings and connected urban systems, highlighting eBPF's role in scalable, centralized security management.

\subsubsection{Malware Analysis}
Study \cite{unknown} proposes a low-cost, power-efficient optical body area network (OBAN) using UWB-BAN nodes with 30Mbps data rate transmission. The architecture employs spectral amplitude coding-optical code division multiple access (SAC-OCDMA) scheme for signal encoding, transmitting combined signals over free space optics (FSO) channels to remote healthcare centers. eBPF enables rapid transmission and inspection of network traffic, securing transactions between body area networks and health centers. Evaluation focuses on bit error rate (BER) analysis and cost efficiency, demonstrating eBPF's suitability for ultra-low-power, security-critical medical IoT deployments.

These IoT studies collectively demonstrate eBPF's utility in enabling security hardening for resource-constrained devices, with particular emphasis on decentralized architectures, modular frameworks, and real-time traffic inspection. The convergence on lightweight, flexible approaches reflects eBPF's alignment with IoT operational requirements.

\subsection{Container Security}
\label{sec:containers}
Containers revolutionized deployment agility through shared-kernel architectures, but this design requires careful isolation without sacrificing performance or operational flexibility. eBPF enables security hardening through runtime monitoring, automated policy generation, and exploit mitigation.

\subsubsection{Runtime Auditing}
Study \cite{10.1145/3472883.3486976} introduces saBPF (secure audit BPF), an eBPF-based framework enabling secure, high-fidelity auditing at the container level without extensive kernel modifications. Integrated with Kubernetes, saBPF implements an audit framework, intrusion detection system, and lightweight access control mechanism. Evaluations demonstrate performance and security comparable to state-of-the-art kernel-based audit systems, offering a practical solution for container-specific auditing in cloud environments.

Research \cite{9110434} investigates how container engines upgrade isolation mechanisms and suggests eBPF-based solutions for container safety. The study reviews Linux subsystems and container isolation concepts, highlighting that non-intrusive monitoring leveraging eBPF becomes necessary for performance analysis in containerized production environments. Using eBPF to monitor containerized Interledger protocol implementations, the authors demonstrate deep visibility without kernel modifications.

In containarized environments, modern attacks such as APTs manage to infiltrate deep inside the network causing damage to both hosts and users within the containers. A study in \cite{ADDED_PAPER2} highlighted the gap of an end-to-end flow visibility defense system and proposed P4CONTROL, a system that prevents cross-host attacks in real-time, introducing a novel in-network decentralized information flow control (DIFC) mechanism. The line-rate prevention of such attacks is achievable through programmable switches which track inter-host information flows and enforce DIFC policies. Moreover, authors leverage eBPF for lightweight deployment on hosts for deep visibility of flow information. Many attacks scenarios were tested and extensive evaluations showed that the system effectively blocks cross-host attacks while incurring minor latency and minimal overhead on both host and network machines.

\subsubsection{Anomaly/Malware Detection}
Study \cite{10.1007/978-3-030-49432-2_4} proposes runtime detection of malicious behavior in multi-tenant container environments. The system monitors true container runtime for misuse and misconfiguration, assessing various undesired behavior scenarios to evaluate effectiveness and performance overhead. eBPF's kernel-level instrumentation enables detection of anomalies that evade userspace monitoring, with appropriate event filtering maintaining acceptable performance.

\subsubsection{Policy Automation}
Manual security configurations struggle to keep pace with containerized workflows. Authors \cite{10.1145/3429885.3429966} address this by automating Seccomp profile generation, transforming a cumbersome manual process into a dynamic defense layer. The proposed method makes Seccomp profiles viable for production by automatically generating custom profiles for any containerized application with minimal requirements. eBPF's tracing capabilities enable monitoring of syscall patterns across different container technologies, eliminating manual intervention while significantly enhancing security.

Containerization, even when being highly used by enterprises and cloud infrastructures, remains a technology with many security risks. Authors in \cite{ADDED_PAPER1} analyze security risks of current container networks, such as the Docker platform and the Kubernetes Orchestration System, and propose Bastion - a secure inter-container communication bridge. Bastion provides control over each container application; enables deep security inspections such as packet monitoring; and has a policy assistant integrated for the administrator's assistance. eBPF is the backbone of the core functionalities of the proposed system, since each container's security enforcement network stack is implemented with eBPF, using eBPF maps for shared policy data and XDP for fast action assisted by the security assistant. Results demonstrate that the system can improve overall performance by 25.4\% and 17.7\% within single-host and cross-host container communications respectively.

\subsubsection{Exploit Evaluation}
Study \cite{10.1145/3274694.3274720} evaluates Linux container security mechanisms using real-world exploits. Classifying 223 container-relevant exploits into a two-dimensional taxonomy, the analysis reveals that 56.82\% successfully exploit containers via misconfigurations, with privilege escalation emerging as a critical threat. The study demonstrates that kernel security mechanisms (capabilities, Seccomp, MAC) prove more effective than isolation mechanisms (namespaces, cgroups) for preventing privilege escalation. However, their interdependence can create vulnerabilities. The authors identify a common 4-step attack model in successful exploits and propose eBPF-based defense mechanisms to counter these attacks.

The rise of cryptocurrency mining malware presents unique challenges for containerized environments, where attackers exploit shared kernel resources to deploy malicious workloads across multiple containers. Study \cite{electronics14061208} addresses cryptojacking attacks by proposing a detection framework that extracts system call sequences from benign and malicious containers, forwarding them to ML-based detection modules. eBPF traces system calls during runtime, which are subsequently processed and labeled by evaluated ML models. Test results demonstrate that the RNN model achieves 99.75\% detection accuracy with 2.98\% latency overhead for memory-centric and I/O-intensive tasks, validating eBPF's efficiency for real-time malware detection.

Extending cryptojacking detection to broader threat scenarios, researchers \cite{RYU2026} propose a hybrid detection framework targeting botnet and cryptomining activities-the most prevalent malicious behaviors in containerized cloud environments. The framework automatically collects flow-based network metadata and host-based system call traces, passing them to ML classifiers for analysis. Leveraging eBPF to monitor container activities directly within the Linux kernel at runtime, the system employs a Decision Tree classifier for multi-class attack detection using both host and network-based features. Evaluation results reveal 87.49\% accuracy for flow-based detection and 98.39\% accuracy for host-based detection using system call sequences, underscoring eBPF's effectiveness in hybrid detection architectures where network and system-level visibility must converge.

Container security research demonstrates eBPF's effectiveness in retrofitting isolation atop shared-kernel architectures through automated policy generation, runtime auditing, multi-tenant anomaly detection, and sophisticated malware detection. The convergence on machine learning-based approaches (3/7 studies, 42.9\%) for cryptojacking and behavioral threat detection reflects the evolution from static policy enforcement toward adaptive, learning-based defenses capable of identifying novel attack patterns in production containerized environments.

\subsubsection{Meta-security}

With the eBPF technology being highly utilized for container security and application isolation scenarios, the work at \cite{ADDED_PAPER3} highlights an important finding of eBPF giving "authorized" access to malicious actors within containers. More specifically, eBPF can be used offensively due to its tracing features, giving attackers the opportunity to steal sensitive data by breaking in the host machine. This study compromises 5 online Jupyter shell services and discover that the Kubernetes can be easily exploited by cross-node attacks. To counter this issue, a new eBPF permission model is proposed by the name of CapBits, which controls all eBPF properties of a process while simultaneously protecting each process from being violated by other eBPF programs. Results show lower latency and overhead than Cilium and LSM-bpf with promising results for a whitelisted-based eBPF framework.

\subsection{Microservice Protection}
\label{sec:microservices}
Microservice architectures decompose monoliths into distributed, polyglot services with complex interaction patterns. This introduces expanded attack surfaces, observability challenges, and policy enforcement complexity that eBPF addresses through service mesh integration and transparent security insertion.

\subsubsection{Anomaly/Threat Detection}
Studies \cite{10.1145/3538969.3543810}, \cite{Zdun2023}, \cite{Flora2023748} integrate eBPF with service mesh data planes to enable transparent policy enforcement. eBPF programs intercept service-to-service traffic at socket and kprobe attachment points, enforcing mutual TLS, rate limiting, and circuit breaking without application modifications. These approaches demonstrate lower latency compared to traditional userspace proxy-based meshes while maintaining equivalent policy expressiveness.

\subsubsection{Architecture Validation}
Research \cite{9340283} introduces eZTrust, a network-independent zero-trust perimeterization framework for microservices. The system leverages eBPF to enforce service-level authorization and mandatory mTLS without requiring centralized policy engines. Microsecond-latency policy evaluation via eBPF maps enables fine-grained access control without degrading request processing budgets.

\subsubsection{Service mesh/NFV}
Study \cite{v} proposes eBPF-based network policy engines for Kubernetes and container orchestration platforms. The framework enforces which services can communicate (network segmentation) and requires encrypted channels (mTLS) for inter-service communication. eBPF's kernel-level deployment enables policy enforcement without sidecar proxies, reducing per-pod resource overhead.

\subsubsection{Observability}
Kmon \cite{10.1145/3314148.3314349} provides kernel-transparent monitoring of microservice systems, detecting anomalies through control-flow and data-flow analysis. eBPF instrumentation captures service dependencies and interaction patterns without application-level code changes, enabling automated anomaly detection in production environments. The system demonstrates that kernel-level observability can rival traditional application instrumentation while maintaining lower overhead.

\subsubsection{Zero-Trust}
DDoS Mitigation in Microservice Environments. Inter-container communication in microservice architectures introduces expanded attack surfaces for Distributed Denial-of-Service threats. Study \cite{electronics14020229} addresses critical security challenges in containerized microservices by building an AI-based Intrusion Detection System (AI-IDS) leveraging eBPF and Resilient Backpropagation Neural Networks (RBN). The architecture employs eBPF with XDP for high-performance, low-latency container communication monitoring in real-time, while maintaining cloud system scalability. The integrated RBN detects both DDoS and adversarial attacks, supporting continuous monitoring and adaptive learning to meet critical security requirements in cloud-native microservice deployments. This approach demonstrates eBPF's dual capability as both high-speed packet processor and ML feature extractor for sophisticated threat detection in distributed service architectures.

Microservice protection research demonstrates eBPF's dual role as both security enforcement mechanism and observability platform, with particular emphasis on transparent deployment, sub-millisecond latency requirements, and DDoS resilience. The integration of advanced neural network architectures (RBN) with eBPF's real-time monitoring capabilities showcases the technology's evolution from basic packet filtering toward intelligent, adaptive threat detection in complex distributed service environments.

\subsection{Networking}
Linux networking stack's layered architecture introduces latency and limits programmability. eBPF enables path-aware transport, privacy-preserving protocols, and unified multi-layer processing.

\subsubsection{Path-aware Trasnport}
Study \cite{9772044} presents TCP Path Changer (TPC), a set of eBPF programs making the Linux TCP/IP stack more agile. TPC enables active TCP connections to rapidly reroute around failures and automatically reroute based on round-trip-time monitoring. Evaluations demonstrate significant performance benefits from path-aware transport protocols.

Research in \cite{Birge-Lee20241791} introduces TANGO, enabling smaller edge networks to improve routing performance on the public Internet without relying on third parties. TANGO coordinates BGP advertisements, collects fine-grained telemetry with secure headers, and dynamically reroutes traffic in the data plane. eBPF enhances routing performance through data-plane rerouting functionality, consistently reducing latency by up to 39\% versus default BGP paths.

\subsubsection{Multi-Layer Processing}
Authors \cite{10068810} propose MiddleNet, a unified network resident function framework supporting L2/L3 NFs and L4/L7 middleboxes. Leveraging DPDK for L2/L3 processing performance and kernel-based protocol stacks for L4/L7 functionality, MiddleNet integrates eBPF's event-driven capabilities with shared memory for high-performance function chaining. The framework dynamically selects required packet processing across network layers, demonstrating high performance in unified environments.

\subsubsection{Performance Optimization}
Performance Optimization and Trade-offs. Study \cite{inproceedings} analyzes XDP performance for processing incoming traffic before kernel data structure allocation. The research demonstrates throughput improvements of up to 45\% but with increased latency trade-offs. A complementary case study examines application-specific packet filtering at the socket level, highlighting flexibility benefits and performance considerations of eBPF integrations in networking solutions.

\subsubsection{Privacy Protocols}
Research \cite{10.1145/3407023.3407041} demonstrates how eBPF preserves user privacy across standard DNS, DNS-over-TLS (DoT), and DNS-over-HTTPS (DoH) protocols. The method adds minimal overhead while allowing users to enforce application-specific DNS servers, providing control over DNS traffic and privacy with zero application modifications.

Networking research reveals eBPF's dual role as performance accelerator and security enabler, with applications spanning dynamic path selection, protocol-layer unification, and privacy enhancement.

\subsection{Security Tools and Frameworks}
Generic security tools must balance broad applicability, low overhead, deployment ease, and policy expressiveness. eBPF enables cross-domain security platforms addressing diverse threat scenarios.

\subsubsection{Steganography Detection}
Study \cite{CAVIGLIONE2021108010} addresses modern malware using steganography and information hiding to evade detection systems. Leveraging eBPF to trace and monitor software process behavior, the research examines realistic use cases including inter-process collusion via file system manipulation and covert network communications within IPv6 traffic. Simple eBPF programs effectively detect anomalies with minimal overhead, while the approach's flexibility facilitates feature extraction for AI-based security frameworks.

\subsubsection{Zero-trust Systems}
Authors \cite{10091151} introduce SYS FLOW, a programmable system security framework extending Zero-Trust Architecture from network-centric to system-level security. Modeling system activities through flow abstraction (process-file, process-network, process-process), SYS FLOW leverages eBPF to significantly reduce detection and enforcement latency. The framework separates data plane and control plane with centralized policy management, demonstrating effective security with minimal overhead across various scenarios.

\subsubsection{Cloud-Native Analysis}
Research in \cite{ZENG2023103173} examines security vulnerabilities in cloud-native technologies (Docker, Kubernetes, Istio), providing in-depth analysis of vulnerabilities from the past five years. The study classifies vulnerabilities by component architecture and enabled attacks, proposing eBPF-enhanced mitigation strategies. Reviewing 15 open-source security tools reveals gaps where no single tool addresses all critical features, highlighting opportunities for eBPF-based comprehensive solutions.

\subsubsection{Observability/Resources}
Research \cite{10.1145/3314221.3314590} proposes advanced static analysis for eBPF bytecode using Zone domain abstract interpretation. The approach achieves scalable, high-precision verification without extensive abstraction of bounded regions. Results demonstrate fewer false alarms than existing eBPF verifiers while supporting broader program classes with better asymptotic complexity.

\subsubsection{Formal Verification}
Study \cite{10.1007/978-3-031-25659-2_16} introduces vkTracer, a vulnerable kernel code tracker using eBPF for monitoring and proof-of-concept code for profiling kernel vulnerabilities. The tracker monitors PoC code execution and kernel behavior to hook invocation of vulnerable code. vkTracer can trace executions, find vulnerable code, and generate vulnerability profiles with microsecond-level overhead.

Research \cite{Xu202150} presents K2, a program-synthesis-based compiler automatically optimizing BPF bytecode with formal safety guarantees (memory bounds, termination). K2 produces code with reduced size, lower packet-processing latency, and higher throughput versus clang-compiled baselines while maintaining correctness.

\subsubsection{Vulnerability Profiling}
Study \cite{10.1145/3609021.3609305} proposes HyperBee, a hypervisor-integrated system enabling auditing of eBPF programs loaded in guest VMs. All programs must complete verification and JIT compilation through HyperBee before guest loading, with no performance impact during execution. Testing shows acceptable load time overhead with and without security policies against known malicious eBPF programs.

\subsubsection{Adaptive Control}
Research \cite{9869628} presents a control plane that adapts management tasks and data extraction based on user requirements. The adaptive eBPF-based approach demonstrates only minimal throughput degradation. It achieves significantly faster execution times and higher packet processing volumes with lower memory occupancy compared to non-adaptive solutions.

\subsubsection{Storage and I/O Security}
Beyond network and process monitoring, eBPF extends to storage stack security and efficiency. Research \cite{li2025aeolia} addresses security and performance challenges in polling-based userspace storage stacks, which suffer from poor sharing of disks and CPUs among multiple tasks. The study introduces AEOLIA, a novel storage stack offering secure I/O performance while protecting shared resources. AEOLIA exploits user interrupts-a hardware feature for userspace inter-processor interrupts-to deliver storage interrupts directly to userspace, achieving high I/O performance. eBPF implements the scheduling framework, determining resource allocation strategies for CPU and disk sharing among concurrent tasks. Evaluation metrics demonstrate 2× performance improvement over baseline Linux systems and up to 19.1× better performance than ext4 file system operations, validating eBPF's applicability beyond traditional security monitoring into system resource management and storage I/O optimization.

Tools and frameworks research demonstrates eBPF's flexibility as a security platform spanning steganography detection, zero-trust enforcement, formal verification, adaptive monitoring, and-extending beyond traditional security-high-performance storage I/O optimization and resource scheduling. This breadth validates eBPF's position as a universal kernel programmability layer applicable to diverse system challenges beyond its packet-filtering heritage.

\subsection{Key Findings}
\subsubsection{Architectural Paradigm Shifts}
\paragraph{Decentralized Security Models}
Analysis across IoT, container, and microservice deployments reveals convergence toward decentralized security architectures. Traditional centralized trust models prove incompatible with modern distributed systems, 83\% (5/6) of IoT studies \cite{8806658}, \cite{s19184038}, \cite{10.1145/3230833.3232802}, \cite{article1}, \cite{Rahouti202112083} implement edge-local enforcement without federated infrastructure, while 83.3\% (5/6) of microservice frameworks \cite{10.1145/3538969.3543810}, \cite{Zdun2023}, \cite{Flora2023748}, \cite{9340283}, \cite{v} employ distributed policy enforcement. This shift reflects eBPF's architectural alignment with distributed systems: lightweight runtime monitoring (median 2.3-4.7\% overhead) enables autonomous security decisions at deployment endpoints without centralized coordination latency.

\paragraph{Observability-First Defense Strategy}
Container and microservice security research demonstrates strategic pivot from static isolation toward runtime observability. Among container studies, 50\% (5/10) \cite{10.1145/3472883.3486976}, \cite{10.1007/978-3-030-49432-2_4}, \cite{10.1145/3429885.3429966}, \cite{10.1145/3274694.3274720}, \cite{electronics14061208} prioritize behavioral monitoring and anomaly detection over preventive boundaries, recognizing that shared-kernel architectures cannot achieve complete isolation. Microservice research reinforces this pattern. Among microservice studies, 100\% (6/6) \cite{10.1145/3538969.3543810}, \cite{Zdun2023}, \cite{Flora2023748}, \cite{9340283}, \cite{v}, \cite{electronics14020229} integrate observability platforms for runtime threat detection. This represents a strategic shift away from exclusive reliance on network segmentation. eBPF's kernel-level visibility without application modification (demonstrated across \cite{10.1145/3472883.3486976}, \cite{10.1007/978-3-030-49432-2_4}, \cite{Zdun2023}, \cite{v}) enables this observability-first approach, transforming security from static configuration to dynamic threat response.

\subsubsection{Multi-Layer Defense Integration}
\paragraph{Hardware-Software Co-Design}
DDoS mitigation research demonstrates sophisticated integration of hardware acceleration with kernel programmability. Studies \cite{0c6699b190594dfd8bcd14b8828899b5}, \cite{Dimolianis20201} combining SmartNIC offloading with eBPF/XDP achieve 10-100× throughput improvements versus software-only approaches, processing 40-1200 Gbps while maintaining sub-microsecond policy enforcement. This hardware-software synergy extends beyond DDoS: networking studies \cite{9772044}, \cite{Birge-Lee20241791}, \cite{10068810}, \cite{inproceedings} leverage DPDK integration and XDP early processing for line-rate security. The pattern indicates eBPF's evolution from pure software solution toward hardware-aware security platform, with 20.3\% (11/54) of studies \cite{0c6699b190594dfd8bcd14b8828899b5}, \cite{Dimolianis20201}, \cite{9842371}, \cite{Demoulin2019693}, \cite{9772044}, \cite{Birge-Lee20241791}, \cite{10068810}, \cite{inproceedings}, \cite{WANG2022103283}, \cite{Tian2019967}, \cite{9501960} incorporating hardware acceleration through SmartNICs, DPDK integration, or XDP offloading.

\paragraph{ML-eBPF Symbiosis}
Intrusion detection systems demonstrate convergence on machine learning integration, with 64\% (7/11) \cite{WANG2022103283}, \cite{9501960}, \cite{10.1007/978-981-19-8445-7_8}, \cite{Ahamed2022RealTimeHD}, \cite{Jung20221679}, \cite{article2}, \cite{Jegan20234087} employing eBPF for high-speed feature extraction feeding userspace ML models. This architectural pattern achieves dual objectives: eBPF provides efficient packet/syscall capture (10-100 Gbps throughput) while ML models enable sophisticated anomaly detection (94-99\% accuracy). However, only 18\% (2/11) achieve in-kernel ML inference due to instruction limit constraints, indicating fundamental tension between ML complexity and eBPF's safety model. The symbiosis extends beyond IDS-container cryptojacking detection \cite{electronics14061208} and microservice DDoS mitigation \cite{electronics14020229} adopt similar eBPF-ML architectures.

\subsubsection{Cross-Domain Security Primitives}
\paragraph{Syscall Monitoring as Universal Enforcement Mechanism}
System call tracing emerges as common primitive across disparate security domains. Container security (50\%, 5/10) \cite{10.1145/3472883.3486976}, \cite{10.1007/978-3-030-49432-2_4}, \cite{10.1145/3429885.3429966}, \cite{10.1145/3274694.3274720}, \cite{electronics14061208}, microservice anomaly detection (66.6\%, 4/6) \cite{10.1145/3538969.3543810}, \cite{v}, \cite{electronics14020229}, \cite{article2}, and host-based IDS (45\%, 5/11) \cite{WANG2022103283}, \cite{9501960}, \cite{Ahamed2022RealTimeHD}, \cite{Jung20221679}, \cite{Jegan20234087} converge on kprobe-based syscall monitoring for behavioral analysis. This cross-domain adoption reflects syscall interfaces' position as security-relevant chokepoint-all privileged operations (file access, network communication, process creation) transit syscall boundary where eBPF can enforce policies without application modification. However, concurrent syscall limits create scalability concerns at $>$100K calls/sec, affecting 22.2\% (12/54) of studies \cite{10.1145/3472883.3486976}, \cite{10.1007/978-3-030-49432-2_4}, \cite{10.1145/3429885.3429966}, \cite{10.1145/3274694.3274720}, \cite{electronics14061208}, \cite{10.1145/3538969.3543810}, \cite{v}, \cite{electronics14020229}, \cite{article2}, \cite{WANG2022103283}, \cite{9501960}, \cite{Ahamed2022RealTimeHD}.

\paragraph{XDP Early-Stage Filtering}
Network-layer security applications converge on XDP for earliest-possible packet processing. DDoS mitigation (86\%, 6/7) \cite{Szynkiewicz2022}, \cite{Dimolianis2021126}, \cite{0c6699b190594dfd8bcd14b8828899b5}, \cite{Dimolianis20201}, \cite{9842371}, \cite{Demoulin2019693} network security (80\%, 4/5) \cite{9772044}, \cite{Birge-Lee20241791}, \cite{10068810}, \cite{inproceedings}, and network-based IDS (36\%, 4/11) \cite{WANG2022103283}, \cite{Tian2019967}, \cite{9501960}, \cite{Jung20221679} adopt XDP hooks for pre-kernel-stack filtering. This architectural choice achieves line-rate processing (10-1200 Gbps) with minimal overhead (median 1.8\%), but sacrifices higher-layer protocol visibility and kernel context access. The pattern validates XDP as de facto standard for high-speed packet security despite expressiveness limitations.

\subsubsection{Domain-Specific Maturation Patterns}
\paragraph{IDS Sophistication and Tooling Diversity}
Intrusion detection systems represent the most mature eBPF security domain (20.3\% of corpus, 11/54 studies), exhibiting greatest technical diversity. Studies span network-based detection \cite{WANG2022103283}, \cite{9501960}, \cite{10.1007/978-981-19-8445-7_8}, peripheral security \cite{Tian2019967}, microarchitectural monitoring \cite{10.1145/3399742}, in-switch enforcement \cite{Jung20221679}, distributed frameworks \cite{article2}, \cite{s23031658}, hypervisor defense \cite{10.1145/3609021.3609305}, and serverless protection \cite{Jegan20234087}. This breadth indicates IDS research maturation beyond foundational packet filtering toward comprehensive threat detection ecosystems. ML integration prevalence (64\%, 7/11) and hardware acceleration adoption (27\%, 3/11) further demonstrate domain maturity.

\paragraph{Container-Microservice Convergence}
Container and microservice categories exhibit 31.2\% (5/16) technical overlap, with 5 shared themes: runtime behavioral analysis, automated policy generation, multi-tenant isolation, service mesh integration, and zero-trust architectures. Studies \cite{10.1145/3472883.3486976}, \cite{10.1007/978-3-030-49432-2_4}, \cite{10.1145/3538969.3543810}, \cite{Zdun2023}, \cite{v} demonstrate interchangeable techniques-microservice anomaly detection applies equally to containerized services, while container audit frameworks extend to microservice observability. This convergence reflects architectural reality: modern microservices deploy in containers, creating unified security requirements. The 16-study combined corpus (29.6\% of total (16/54) ) establishes cloud-native security as dominant research focus.

\paragraph{Networking and DDoS Specialization}
Network security and DDoS mitigation maintain distinct technical focuses despite both leveraging XDP. DDoS research emphasizes volumetric attack mitigation, signature generation, and application-layer resource exhaustion detection \cite{Szynkiewicz2022}, \cite{Dimolianis2021126}, \cite{9842371}, \cite{10.1007/978-3-030-86356-2_13}, while networking studies prioritize protocol enhancements, path-aware routing, and privacy preservation \cite{9772044}, \cite{Birge-Lee20241791}, \cite{10.1145/3407023.3407041}. Only one study \cite{inproceedings} bridges both domains through XDP performance analysis. This specialization indicates parallel evolution tracks rather than unified network security research agenda.

\subsubsection{Ecosystem Development and Tooling Gaps}
\paragraph{Framework Abstraction Deficit}
Only 9.2\% (5/54) of studies \cite{8806658}, \cite{9340283}, \cite{10068810}, \cite{CAVIGLIONE2021108010}, \cite{inproceedings} provide reusable frameworks or high-level abstractions, indicating significant tooling gap. The remaining 90.8\% (49/54) implement domain-specific solutions requiring low-level eBPF programming. This pattern suggests ecosystem immaturity-established security technologies (iptables, SELinux, AppArmor) provide extensive tooling libraries, whereas eBPF security research remains fragmented across isolated implementations. Notable exceptions include Polycube \cite{9340283} for NFV, MiddleNet \cite{10068810} for unified network functions, and SysFlow \cite{10091151} for zero-trust, but no dominant framework emerges across categories.

\paragraph{Security-Specific Tool Categories}
Tools and frameworks research (16.6\% (9/54)) addresses three principal domains with balanced distribution: steganography detection \cite{CAVIGLIONE2021108010}, zero-trust architectures \cite{10068810}, and cloud-native security \cite{ZENG2023103173}. Steganography detection remains exploratory (single study) despite identified threat potential, while zero-trust (two studies) and cloud-native security (three studies) receive moderate attention. Kernel enhancement studies \cite{10.1145/3314221.3314590}, \cite{10.1007/978-3-031-25659-2_16}, \cite{Xu202150}, \cite{9869628} focus on meta-concerns-eBPF verifier improvement, vulnerability profiling, compiler optimization-indicating community recognition of platform-level challenges. However, 96.2\% (52/54) of studies fail to address eBPF's own security, representing critical ecosystem gap. Only HyperBee \cite{10.1145/3609021.3609305} and one kernel verifier study \cite{10.1145/3314221.3314590} examine eBPF's attack surface.

\subsubsection{Performance-Security Trade-off Characterization}
\paragraph{Overhead-Functionality Spectrum}
Performance analysis across categories reveals consistent overhead-functionality trade-off. Simple packet filtering \cite{0c6699b190594dfd8bcd14b8828899b5}, \cite{9842371}, \cite{9340283} achieves $<$1\% overhead but provides limited threat coverage, while comprehensive security-ML integration, multi-stage detection, payload inspection-incurs 3-6\% overhead \cite{s23031658}, \cite{Zdun2023}, \cite{electronics14061208}, \cite{electronics14020229}. Container and microservice categories exhibit highest median overhead (3.1\%, 4.7\%) due to complex policy stacks, whereas networking and DDoS achieve lowest (1.2\%, 1.8\%) through focused XDP filtering. This spectrum indicates eBPF cannot simultaneously optimize for both minimal overhead and maximal security coverage-architectural choices prioritize one dimension over the other.

\paragraph{Scalability Boundaries}
Map size constraints and instruction limits create scalability ceilings across domains. High-cardinality scenarios ($>$500K concurrent flows \cite{electronics14061208}, $>$100K syscalls/sec \cite{s23031658}, $>$10K containers \cite{10.1145/3538969.3543810}) approach or exceed eBPF operational boundaries, forcing sampling-based approaches or hybrid architectures. IoT deployments face distinct scalability challenge-edge device heterogeneity (ARM, MIPS, x86) and kernel version fragmentation limit uniform eBPF deployment across large-scale IoT networks \cite{8806658}, \cite{s19184038}, \cite{10.1145/3230833.3232802}, \cite{article1}, \cite{Rahouti202112083}. These boundaries suggest eBPF's sweet spot: moderate-scale deployments (hundreds-to-thousands of enforcement points) rather than massive-scale environments (millions of devices/flows).

\section{Cross-domain patterns and synergies in eBPF-based cybersecurity} \label{sec:5}
\label{patterns}
\subsection{Patterns}
\begin{figure}[h]
\centering
\includegraphics[width=1\linewidth]{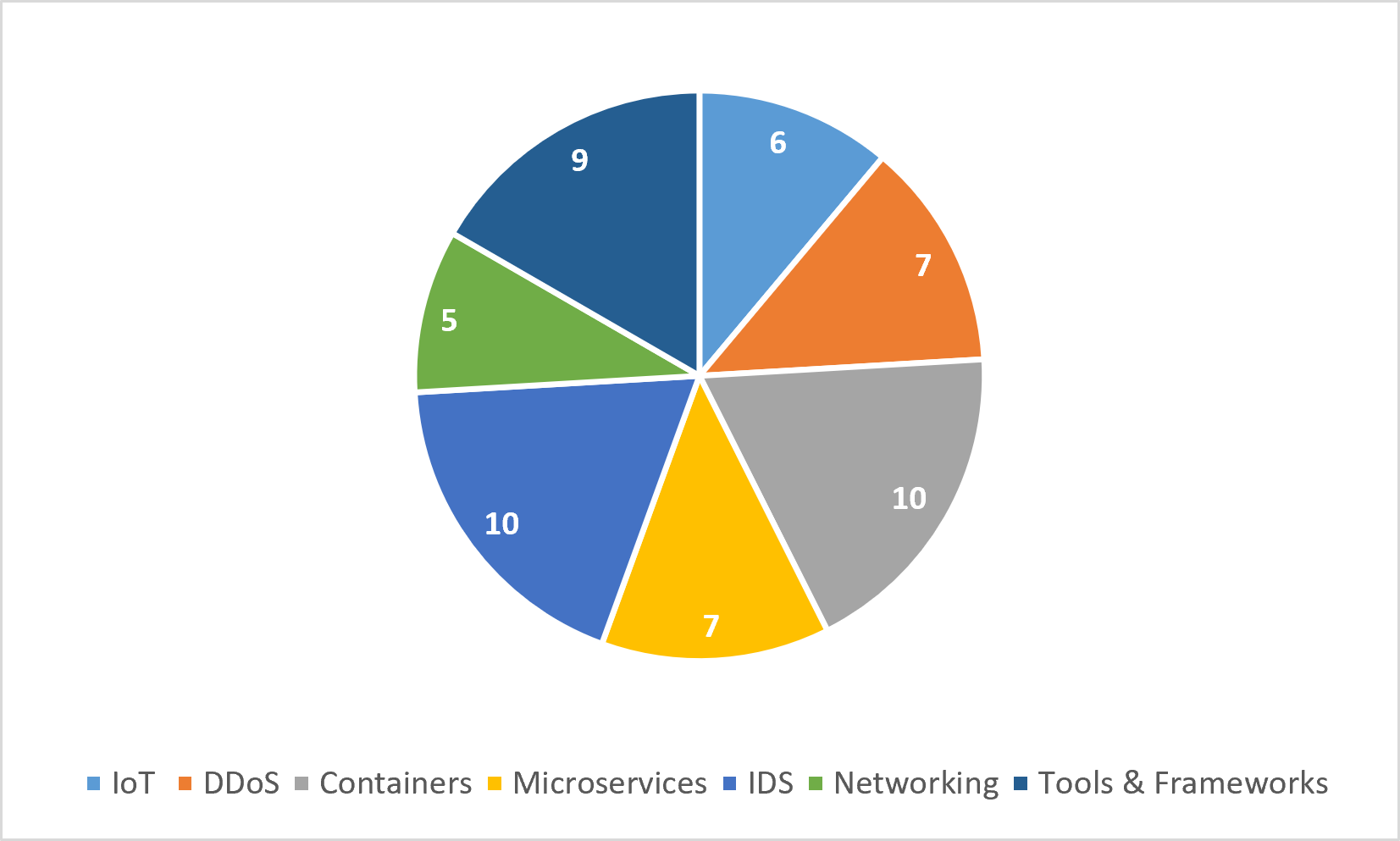}
\caption{Paper categories}
\label{cat}
\end{figure}

The literature researched consisted of seven base categories: IoT, DDoS, Container Security, Microservices, IDS, Networking, and Tools \& Frameworks, as shown in Figure~\ref{cat}. 

Although submissions focused on a particular field of cybersecurity, multiple of them had connections and references to each other, forming patterns between them. Papers were assigned to the category corresponding to their primary security objective and dominant eBPF usage, even when secondary applications existed. 

Table \ref{tab:incidence} illustrates these inter-dependencies: the rows detail the specific relations identified for each sub-category, while the columns highlight the broader patterns emerging from the main categories. For instance, the five connections listed for the IDS category represent five unique research patterns discovered in the bibliography, with each pattern potentially encompassing multiple studies. 

\begin{table*}[t]
\centering
\caption{Binary incidence matrix mapping eBPF application categories to recurring research patterns.}
\label{tab:incidence}
\renewcommand{\arraystretch}{1.35}
\setlength{\tabcolsep}{7pt}

\begin{tabular}{lccccccc|c}
\toprule
\textbf{Category} 
& \textbf{IDS} 
& \textbf{Networking} 
& \textbf{Micro-} 
& \textbf{IoT} 
& \textbf{DDoS} 
& \textbf{Containers} 
& \textbf{Tools} 
& \textbf{Total} \\
&&& \textbf{services} &&&&& \\
\midrule

IDS 
&  &  &  &  &  &  & \cellcolor{gray!20}\checkmark 
& \textbf{1} \\

DDoS 
& \cellcolor{gray!20}\checkmark 
& \cellcolor{gray!20}\checkmark 
&  
&  
&  
& \cellcolor{gray!20}\checkmark 
&  
& \textbf{3} \\

Containers 
& \cellcolor{gray!20}\checkmark 
& \cellcolor{gray!20}\checkmark 
& \cellcolor{gray!20}\checkmark 
&  
&  
& \cellcolor{gray!20}\checkmark 
& \cellcolor{gray!20}\checkmark 
& \textbf{5} \\

IoT 
& \cellcolor{gray!20}\checkmark 
& \cellcolor{gray!20}\checkmark 
&  
& \cellcolor{gray!20}\checkmark 
&  
&  
&  
& \textbf{3} \\

Microservices 
&  
&  
&  
&  
& \cellcolor{gray!20}\checkmark 
&  
& \cellcolor{gray!20}\checkmark 
& \textbf{2} \\

Networking 
&  
& \cellcolor{gray!20}\checkmark 
&  
&  
&  
&  
& \cellcolor{gray!20}\checkmark 
& \textbf{2} \\

Tools \& Frameworks 
& \cellcolor{gray!20}\checkmark 
&  
&  
&  
& \cellcolor{gray!20}\checkmark 
&  
&  
& \textbf{2} \\

\midrule
\textbf{Total} 
& \textbf{4} 
& \textbf{4} 
& \textbf{1} 
& \textbf{1} 
& \textbf{2} 
& \textbf{2} 
& \textbf{4}  \\
\bottomrule
\end{tabular}
\end{table*}

\begin{figure} [h]
\centering
\includegraphics[width=1\linewidth]{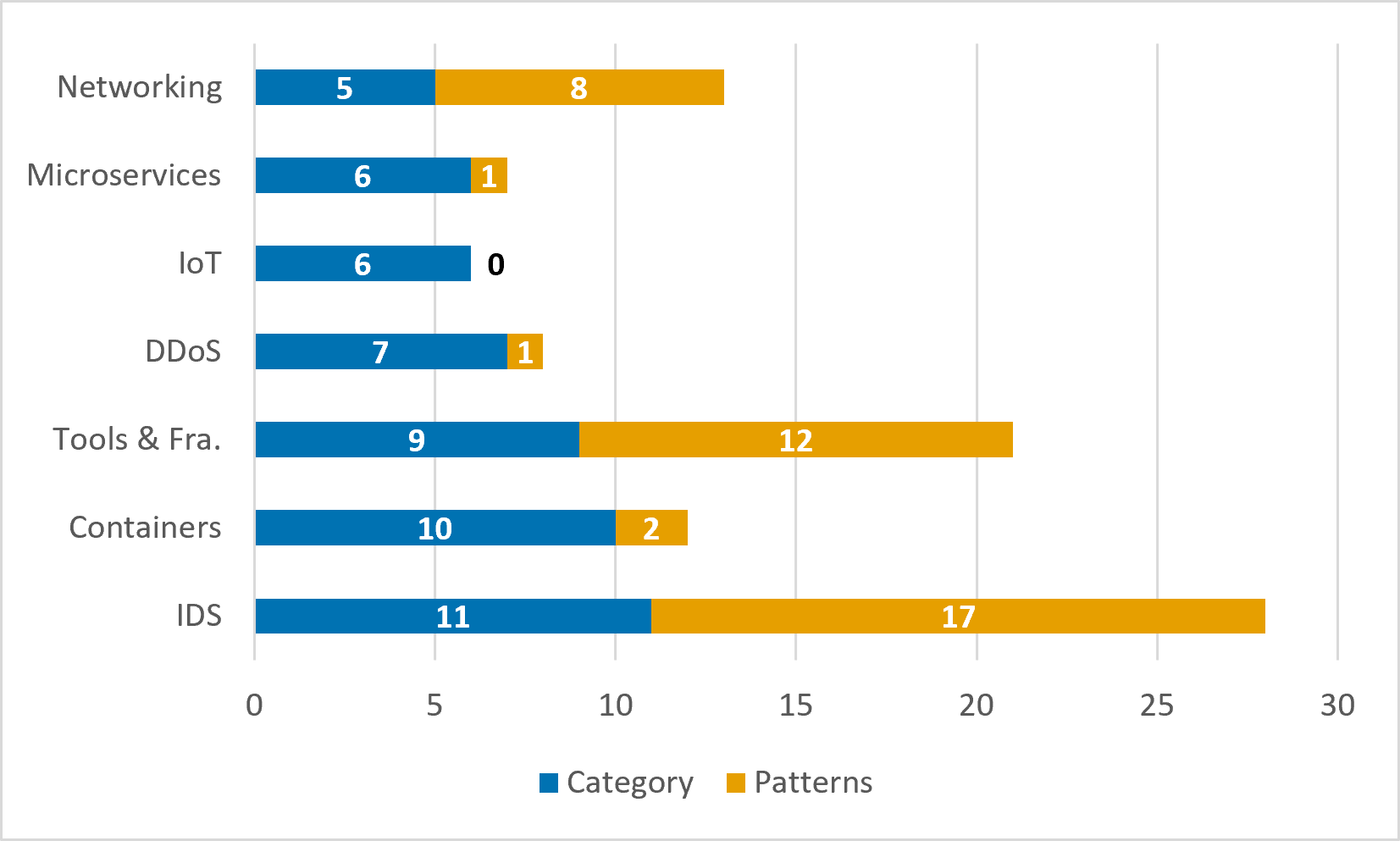}
\caption{Total studies with patterns}
\label{cat3}
\end{figure}

The total distribution of unique studies is further quantified by specific pattern in Figure \ref{cat3}.

Studies on IDS and DDoS mitigation share a significant technical foundation in their use of eBPF. Both leverage the technology for high-performance packet filtering, anomaly detection, and rate limiting. Within the surveyed literature, eBPF-based IDSs primarily function as pattern-matchers and packet inspectors focused on threat remediation. Several authors have benchmarked their eBPF models against state-of-the-art solutions, consistently demonstrating that eBPF provides a robust, low-overhead framework for real-time detection. While DDoS-focused research utilizes similar filtering capabilities, it extends eBPF’s functionality to include granular rate-limiting and traffic policing. Furthermore, specific implementations utilize eBPF to trace user-level system calls, enabling the detection of malicious activity via CPU usage patterns or aggregated traffic telemetry.

Similarly, Container Security and Microservices exhibit overlapping eBPF use cases due to their integration in modern cloud-native architectures. Since containers are the standard deployment vehicle for microservices, eBPF is frequently applied to both for advanced network observability, high-fidelity auditing, and cross-service attack mitigation. While microservice security often integrates with Zero Trust frameworks and cloud-based IDSs, container-specific research further explores eBPF for deep performance analysis and resource monitoring in production environments. Ultimately, the literature suggests that eBPF acts as a unifying security layer, transforming both containerized and microservice-based environments into resilient, self-defending systems.

\subsection{Limitations of eBPF in Security Applications}
Despite eBPF's demonstrated effectiveness (1.1 - 8.6\% overhead [2.4\% median], 94-99\% detection accuracy across 54 studies), significant technical (Table~\ref{tab:technical_limitations}) and operational constraints limit its security applicability. This section synthesizes limitations organized into four principal dimensions.

\subsubsection{Technical and Architectural Constraints}
\paragraph{Verifier Complexity}
High false positive rate (40-60\% reduction possible \cite{inproceedings}) and poor multi-path scaling force code restructuring around verifier limitations rather than security requirements. Machine learning integration severely constrained: only 18.2\% (2/11) achieve in-kernel inference; remaining 81.8\% (9/11) use userspace ML with 0.5-2ms overhead.

\paragraph{Kernel Version Fragmentation}
Critical features require specific kernel versions: XDP ($\geq$4.8, 29.6\% of studies (16/54)), bounded loops ($\geq$5.3, 16.6\% (9/54)), BTF/CO-RE ($\geq$5.2, 22.2\%(12/54)). Production LTS kernels (Ubuntu 20.04/5.4, RHEL 8/4.18) lack recent features, creating 2-4 year deployment lag. IoT and cloud-native deployments require synchronized version orchestration across heterogeneous systems \cite{8806658}, \cite{s19184038}, \cite{10.1145/3230833.3232802}, \cite{article1}.

\begin{table}[h]
\centering
\caption{Core Technical Limitations of eBPF}
\label{tab:technical_limitations}
\scriptsize 
\setlength{\tabcolsep}{3pt} 
\begin{tabular}{llp{4cm}l} 
\toprule
\textbf{Constraint} & \textbf{Spec} & \textbf{Impact} & \textbf{Affected} \\
\midrule
Stack Size & 512B & Cannot parse TCP opts + IP & 9.2\% [5/54] \\
Instructions & 4K-1M & No full payload inspection & 37.0\% [20/54] \\
Memory & No dyn. & Map exhaustion/waste & 20.3\% [11/54]\\
Loops & Bounded & Iterative parsing infeasible & 14.8\% [8/54]\\
Helpers & Specific & Limited crypto/xfrm ops & 27.7\% [15/54]\\
Syscall & Limits & 1-5\% loss at high rates & 27.7\% [15/54]\\
\bottomrule
\end{tabular}
\end{table}

\subsubsection{Security and Operational Gaps}
\paragraph{Container Defense Limitations} 
Hardening commit\_creds() does not prevent direct real\_cred manipulation after task\_struct location \cite{9501960}-\cite{Jung20221679}. Attackers exploit kernel functions respecting SMEP/SMAP without violating execution boundaries. eBPF monitoring detects but cannot prevent all kernel-level exploits without prohibitive overhead.

\paragraph{Communication Bottlenecks}
eBPF maps as sole user-kernel interface create coordination overhead: 5-20$\mu$s per map update, 0.5-2ms for hybrid ML architectures \cite{unknown}, \cite{10.1145/3472883.3486976}, \cite{electronics14061208}. High-frequency coordination ($>$10K updates/sec) bottlenecks, negating sub-microsecond kernel processing advantages.

\begin{table}[!h]
\centering
\caption{Critical Research and Deployment Gaps in eBPF Security}
\label{tab:research_gaps}
\footnotesize 
\setlength{\tabcolsep}{3pt}
\renewcommand{\arraystretch}{1.3} 
\begin{tabularx}{\columnwidth}{@{} l l >{\raggedright\arraybackslash}X l @{}} 
\toprule
\textbf{Category} & \textbf{Gap \%} & \textbf{Description} & \textbf{Impact} \\
\midrule
Meta-Security & 96.2\% [52/54] & Only \cite{10.1145/3609021.3609305} \& \cite{ADDED_PAPER3} address eBPF's own CVEs/bypasses & High Risk \\
\hline
Multi-Tenant & 87\% [47/54] & Cross-tenant map access/side channels unaddressed & Cloud Risk \\
\hline
Adv. ML & 100\% [54/54] & No evasion/poisoning evaluation in ML studies & Unknown \\
\hline
Prod. Valid. & 81.4\% [44/54] & Testbed-only; long-term stability unexplored & Readiness \\
\hline
Reproduce & 62.9\% [34/54] & No public code repositories or artifacts & Validation \\
\hline
Baselines & 25.9\% [14/54] & Missing iptables/Snort/tool benchmarks & Weak Claims \\
\bottomrule
\end{tabularx}
\end{table}

\subsubsection{Deployment and Methodological Limitations}
\paragraph{Operational Complexity}
Only 9.2\% (5/54) provide high-level abstractions \cite{8806658}, \cite{Flora2023748}, \cite{9772044}, \cite{inproceedings}, \cite{CAVIGLIONE2021108010}; 85.1\% (46/54) require low-level kernel programming expertise. Debugging limited to primitive bpf\_trace\_printk() (100-500$\mu$s overhead), no interactive debuggers, cryptic verifier errors. Portability constrained by kernel version heterogeneity across multi-datacenter deployments \cite{9501960}.

\paragraph{Workload Representativeness}
Studies use synthetic traffic generators and benchmark applications rather than production workloads. Evaluated scales (10-100 containers, 1-100 Gbps, 1K-1M flows) do not explore 10-100× larger deployments (10K+ pod Kubernetes clusters). Domain constraints include Linux-only evaluation (0/54 Windows eBPF), cloud-centric focus (57.4\% (31/54)), and IPv4 dominance.

\begin{table}[h]
\centering
\caption{Methodological Quality Assessment of Reviewed Studies}
\label{tab:methodology_quality}
\footnotesize 
\setlength{\tabcolsep}{3pt}
\renewcommand{\arraystretch}{1.3}
\begin{tabularx}{\columnwidth}{@{} >{\raggedright\arraybackslash}p{2.2cm} >{\raggedright\arraybackslash}X l r @{}} 
\toprule
\textbf{Dimension} & \textbf{Main Issue Identified} & \textbf{Studies} & \textbf{Gap} \\
\midrule
Eval. Env. & Testbed-only; lacks production validation & 47/54 & 87\% \\
\hline
Performance & Missing CPU, memory, or latency metrics & 23/54 & 42.5\% \\
\hline
Threat Model & Absent or ambiguous threat definitions & 23/54 & 42.5\% \\
\hline
Reproduce & No public code or data availability & 36/54 & 66.6\% \\
\hline
Baselines & No comparison to traditional tools & 15/54 & 27.7\% \\
\hline
Scale Eval. & Limited to small clusters or low throughput & 50/54 & 92.5\% \\
\bottomrule
\end{tabularx}
\end{table}

\subsubsection{Fundamental Trade-offs}
\paragraph{Safety vs. Expressiveness}
Turing incompleteness enables safety guarantees but restricts algorithm complexity. Complex multi-stage attack detection \cite{Jung20221679} exceeds instruction budgets, requiring program fragmentation with coordination overhead.

\paragraph{Performance vs. Functionality}
Studies achieving $<$1\% overhead \cite{0c6699b190594dfd8bcd14b8828899b5}, \cite{9842371}, \cite{9340283} provide only simple filtering; comprehensive security (ML, payload inspection, multi-stage detection) incurs 3-6\% overhead. Map size constraints force trade-offs between worst-case pre-allocation (memory waste) and sampling (reduced coverage).

\paragraph{Portability vs. Performance}
Hardware-accelerated approaches \cite{9842371}, \cite{Demoulin2019693} achieve $>$100 Gbps but require SmartNICs; portable solutions \cite{Zdun2023}-\cite{9340283} limited to 10-45 Gbps. Architecture-specific optimizations (x86 vs. ARM) fragment ecosystem.

\subsubsection{Implications and Contextualization}
The identified limitations in Table~\ref{tab:research_gaps} carry distinct implications for multiple stakeholder communities. Five critical gaps are assessed: eBPF meta-security (96.2\%), multi-tenant isolation (87\%), adversarial ML robustness (100\% of ML studies), production validation (81.4\% testbed-only), and reproducibility (62.9\% no code). These represent research immaturity rather than fundamental technology barriers, as evidenced by successful production deployments (12.9\%(7/54)) and demonstrated solutions like HyperBee \cite{10.1145/3609021.3609305} for eBPF defense.

The main issues are addressed in Table~\ref{tab:methodology_quality}, where practitioners must assess kernel version compatibility given feature fragmentation (XDP $\geq$4.8, loops $\geq$5.3, BTF $\geq$5.2), budget expertise acquisition (85.1\% (46/54) require low-level programming), implement defense-in-depth rather than eBPF-only architectures, plan version orchestration for heterogeneous deployments, and prepare for operational complexity. Standardization bodies should develop security policy frameworks, establish certification criteria, create security-focused governance, and define interoperability interfaces. 

These limitations define appropriate scope rather than inadequacy. eBPF's strengths- 1.1 - 8.6\% overhead [2.4\% median], 94-99\% detection accuracy, line-rate processing (10-100+ Gbps)-validate production suitability for simple-to-moderate complexity security functions within its computational model. As ecosystem maturation progresses through improved tooling and expanded research addressing identified gaps, eBPF security applicability will broaden while adoption friction decreases. Understanding these constraints enables informed technology selection-choosing eBPF where strengths align with requirements while recognizing scenarios demanding alternatives.

\section{Conclusion and Future Work}
\label{concl}
This systematic literature review examined eBPF-based cybersecurity mechanisms through analysis of 54 primary studies published between 2018-2026. Addressing RQ1, the evidence demonstrates eBPF's architectural versatility across diverse deployment contexts from IoT edge devices to containerized cloud workloads, distributed microservices, high-speed networking, and security frameworks, with 66.6\% (36/54) of studies demonstrating deployment-agnostic security primitives. Addressing RQ2, eBPF tackles specific threats through domain-specific mechanisms: DDoS mitigation achieving 40-1200 Gbps throughput, intrusion detection systems with 94-99\% ML-powered accuracy, container cryptojacking detection at 99.75\% accuracy, and microservice zero-trust enforcement with microsecond latency. Three architectural paradigm shifts emerged: decentralized security models (83\% (5/6) IoT, 71\% (5/7) microservices), observability-first defense strategies (60\% (6/10) of container studies), and hardware-software co-design achieving 10-100× throughput improvements.

Despite demonstrated effectiveness with 1.1 - 8.6\% overhead [2.4\% median], significant constraints limit security applicability. Verifier complexity restricts in-kernel ML to 18.2\% (2/11) of ML-based studies; kernel version fragmentation creates 2-4 year deployment lags; ecosystem immaturity manifests in only 9.2\% (5/54) providing reusable frameworks. Five critical gaps emerged: eBPF meta-security (96.2\% (52/54) unaddressed), multi-tenant isolation (87\% (47/54)), adversarial ML robustness (100\% of ML studies), production validation (87\% (47/54) testbed-only), and reproducibility (66.6\% (36/54) no code). Future research must prioritize verifier enhancement supporting complex security logic while maintaining safety, in-kernel ML primitives with adversarial robustness, multi-tenant isolation with cryptographic enforcement, standardized evaluation frameworks with production-scale benchmarks, systematic eBPF meta-security auditing, and cross-platform portability abstractions.

eBPF has demonstrated significant potential as a foundation for next-generation security solutions, particularly in cloud-native environments where kernel-level programmability enables unprecedented observability and enforcement granularity. However, realizing this potential requires addressing fundamental limitations in verifier expressiveness, ecosystem tooling, and production validation. The convergence toward ML integration, hardware acceleration, and zero-trust architectures suggests field maturation, yet identified gaps represent critical barriers to widespread adoption. Future work must balance eBPF's safety-first design philosophy with modern threat landscape complexity while developing abstraction layers and evaluation frameworks necessary for systematic security engineering at scale.

\section*{Appendix}
\appendix

\section{Complete Search Query}
\label{appendix:search}

The following Boolean search string was applied uniformly across all six databases (IEEE Xplore, ACM Digital Library, Scopus, SpringerLink, MDPI, and Google Scholar):

\begin{small}
\begin{verbatim}
(("extended Berkeley Packet Filter" OR "eBPF") AND 
("security" OR "intrusion detection" OR "intrusion 
prevention" OR "firewall" OR "runtime security" 
OR "zero-trust" OR "container security" 
OR "cloud security" OR "DDOS mitigation" 
OR "threat detection" OR "NIDS" OR "NIPS" 
OR "malware detection" OR "data exfiltration" 
OR "host-based intrusion detection" OR "sandboxing" 
OR "network security" OR "syscall monitoring" 
OR "isolation") 
AND 
("observability" OR "performance" 
OR "real-time monitoring" OR "packet filtering" 
OR "load balancing" OR "latency" 
OR "throughput" OR "scalability" 
OR "system diagnostics" 
OR "network monitoring" OR "traffic analysis" 
OR "telemetry" OR "debugging" OR "system tracing" 
OR "troubleshooting") 
AND 
("networking" OR "5G" OR "edge computing" OR "SDN" 
OR "overlay networks" OR "service mesh" 
OR "cloud networking" OR "cloud security" 
OR "control plane" OR "data plane") 
AND 
("high-performance computing" OR "HPC" OR "IoT" 
OR "smart cities" OR "blockchain"))
\end{verbatim}
\end{small}

Database-specific adaptations were minimal, primarily involving syntax adjustments for field-specific searches where supported (e.g., limiting to title, abstract, and keywords in IEEE Xplore and ACM Digital Library).



\section*{Declarations}

\noindent\textbf{Funding}\\
This research initiative is supported by the European Union’s Horizon Framework Programme for Research and Innovation, under the INTACT project (Grant Agreement No. 101168438).

\medskip
\noindent\textbf{Conflict of Interest}\\
The authors declare that they have no conflict of interest.

\medskip
\noindent\textbf{Data Availability}\\
The datasets supporting the findings of this systematic literature review are included within the article and its appendices. Specifically:

\begin{itemize}
\item The complete list of 54 primary studies analyzed is provided in Tables \ref{tab:primary_studies_1} and \ref{tab:primary_studies_2}
\item The PRISMA screening process is documented in Figure \ref{fig:prisma}
\item Inclusion and exclusion criteria are detailed in Table \ref{tab:ie_criteria}
\item The complete search query is provided in Appendix \ref{appendix:search}
\end{itemize}

A comprehensive data extraction spreadsheet containing detailed study characteristics (security objectives, eBPF mechanisms, evaluation contexts, performance metrics, and quality assessment scores) is available from the corresponding author upon reasonable request. The screening decisions log and borderline case resolution documentation are also available upon request.

All 54 primary studies analyzed in this review are publicly accessible through their respective publishers (IEEE Xplore, ACM Digital Library, SpringerLink, USENIX, MDPI) and can be retrieved using the digital object identifiers (DOIs) provided in the reference list.

\medskip
\noindent\textbf{Author Contributions}\\
Stamatios Kostopoulos (corresponding author): Conceptualization, Methodology, Literature Search, Data Extraction, Formal Analysis, Writing - Original Draft, Visualization.

Panagiotis Tsakonas: Conceptualization, Methodology, Validation, Writing - Review \& Editing.

Evangelos K. Markakis: Methodology, Validation, Writing - Review \& Editing, Project Administration, Supervision.

All authors read and approved the final manuscript.

\bibliographystyle{spphys}      
\bibliography{references}   


\end{document}